\documentclass{cta-author}

\usepackage[utf8]{inputenc}
\usepackage{tabularx, makecell, booktabs}

{}
{}
{}
\usepackage{makecell}
\usepackage{siunitx}
\usepackage{amsmath}
\usepackage{breqn}
\usepackage{hyperref}
\usepackage{soul}
\usepackage{graphicx}
\usepackage{url}
\usepackage{enumitem}
\usepackage[style=base]{caption}
\usepackage{amsmath,amssymb,amsfonts}
\usepackage{algorithmic}
\usepackage{comment}
\usepackage[ruled,vlined]{algorithm2e}
\usepackage{comment}
\usepackage{setspace}
\usepackage[outdir=./]{epstopdf}

\begin{document}

\title{A Survey of Machine Learning Methods for Detecting False Data Injection Attacks in Power Systems}
\author{\au{Ali Sayghe$^{1}$}, \au{Yaodan Hu$^2$}, \au{Ioannis Zografopoulos$^1$}, \au{XiaoRui Liu$^1$}, \au{Raj Gautam Dutta$^2$}, \au{Yier Jin$^2$}, \au{Charalambos Konstantinou$^1$}}

\address{\add{1}{FAMU-FSU College of Engineering, Center for Advanced Power Systems, Florida State University, Tallahassee, FL, USA}
\add{2}{Department of Electrical and Computer Engineering,University of Florida, Gainesville, FL, USA}
\email{aas17g@my.fsu.edu, cindy.hu@ufl.edu, izografopoulos@fsu.edu, xliu9@fsu.edu, r.dutta@ufl.edu, yier.jin@ece.ufl.edu, ckonstantinou@fsu.edu}}

\begin{abstract}
Over the last decade, the number of cyberattacks targeting power systems and causing physical and economic damages has increased rapidly. {Among them, False Data Injection Attacks (FDIAs) is a class of cyberattacks against power grid monitoring systems.} {Adversaries can successfully perform FDIAs in order to manipulate the power system State Estimation (SE) by compromising sensors or modifying system data.} SE is an essential process performed by the Energy Management System (EMS) towards estimating unknown state variables based on system redundant measurements and network topology. {SE routines include  Bad Data Detection (BDD) algorithms to eliminate errors from the acquired measurements, e.g., in case of sensor failures.} 
FDIAs can bypass BDD modules to inject malicious data vectors into a subset of measurements without being detected, and thus manipulate the results of the SE process. {In order to overcome the limitations of traditional residual-based BDD approaches, data-driven solutions based on machine learning algorithms have been widely adopted for detecting malicious manipulation of sensor data due to their fast execution times and accurate results. This paper provides a comprehensive review of the most up-to-date machine learning methods for detecting FDIAs against power system SE algorithms.}

\end{abstract}

\maketitle

\section{Introduction}
\label{sec:introduction}
{The first practical power system, developed by Westinghouse Electric company in 1886, changed the landscape of human society ~\cite{kline2005competing}}. {Recently, the integration of Information and Communication Technology (ICT) into power grid applications has enabled the evolution towards a smart grid architecture. Smart grids, among others, improve the monitoring capabilities of power systems leveraging advanced microprocessor-based components such as Phasor Measurement Units (PMUs) and smart meters.} {Grid operators can impose controls on the electricity generation and consumption, increasing the efficiency and reliability of power systems by utilizing the measurements from these components. At the same time, the inclusion of smart sensing and control devices expanded the attack landscape \cite{mclaughlin2016cybersecurity}.}  {The increasing network interfaces of smart grid implementations provide entry points for cyber-intruders \cite{hahn2011cyber}}. In December 2015, a cyberattack on the Ukrainian power grid led to a power outage affecting more than $200,000$ customers ~\cite{ukraine2015}. One year later, a similar but more complex attack was carried out again in Ukraine ~\cite{ukraine2016}. {These attack incidents confirm that the vulnerabilities within grid devices and networks could be maliciously exploited (even remotely) with large-scale impacts on the system \cite{liu2019reinforcement, liu2020deep, keliris2019open}.} 

It is critical to detect cyberattacks promptly to increase the security and reliability of the power system.  {This paper focuses on False Data Injection Attacks (FDIAs), a type of cyberattacks that inject false measurements to poison the State Estimation (SE) process \cite{shweppe1969power}.}  {Traditional Bad Data Detection (BDD) methods are based on the residuals between the observed and estimated measurements ~\cite{bandak2014power, monticelli1983reliable, zakerian2017bad}; if the residual is larger than a threshold, bad data is suspected to exist. Despite the wide adoption of such methods, it has been demonstrated that FDIAs can bypass BDD algorithms.
{The concept of FDIAs in power systems was introduced in 2009~\cite{liu2011false}. Different techniques have been proposed since then to detect FDIAs including the Kullback Leibler distance method, fast Go-Decomposition, Unscented Kalman Filter (UKF), Bayesian formulation, Bayesian framework, generalized likelihood ratio, Markov chains, cosine similarity matching scheme and {diagnostic robust generalized potential} \cite{ musleh2020survey, zhang2019false, aoufi2020survey, majumdar2016bad, boba, liu2014detecting, chaojun2015detecting, li2018detecting, vzivkovic2018detection, kosut2010limiting, kosut2010malicious, li2015quickest, rawat2015detection, singh2018joint}.}} However, such techniques often fail to detect FDIAs that fit the same distribution of historical measurements and can only capture attacks that cause abnormal system states~\cite{singh2018joint}. {For example, the Kullback Leibler distance method fails to detect FDIAs in system buses where the attacker injects a small measurement error into a specific state.} Also, the Bayesian framework and generalized likelihood ratio methods cannot detect FDIAs if the attacker replaces the current meter readings with historical readings that have the same distribution. {To address this issue, Majumdar \textit{et al}. proposed a technique called diagnostic robust generalized potential \cite{majumdar2016bad}. First, the system measurements are separated in leverage and non-leverage sets, and then by employing the diagnostic robust generalized potential method, bad data can be efficiently identified performing residual analysis, even if FDIAs exist in the form of gross errors. However, it is well-known that identifying bad leverage points is challenging for such largest normalized residual statistical tests \cite{zhao2018vulnerability}. }

Machine learning algorithms have been widely applied in power grid functions for control and monitoring purposes~\cite{zhang2011distributed, anderson2011adaptive, rudin2012machine}. For example, Zhang \textit{et al}. implemented analyzing modules leveraging machine learning algorithms at different levels of the grid network for intrusion detection~\cite{zhang2011distributed}. Anderson \textit{et al}. proposed a machine learning algorithm to manage the system loads and sources ~\cite{anderson2011adaptive}. Rudin \textit{et al}. suggested using machine learning algorithms to anticipate component failures in power systems~\cite{rudin2012machine}. 
{In order to overcome FDIAs detection limitations, researchers have also developed techniques leveraging machine learning algorithms to efficiently detect such attacks~\cite{ozay2016machine, esmalifalak2011stealth, svm2, musleh2020survey, aoufi2020survey}. Various types of algorithms have been investigated in literature including supervised, semi-supervised, unsupervised, and deep learning. Such methods demonstrate better performance in terms of accuracy and adaptability to dynamic and uncertain grid environments~\cite{wilson2018deep, yan2016detection, ozay2016machine, esmalifalak2011stealth, svm2}. }

In this work, we present a survey of FDIAs detection methods based on machine learning algorithms. The contributions of this paper are as follows:
\begin{itemize}[leftmargin=3em,labelsep=1em]
    \item We present a comprehensive overview of FDIAs in the power grid including background information for SE, different FDIAs settings, impacts of FDIAs on power systems, and FDIAs defense methods.
    \item {We provide a survey of FDIAs detection methods based on the machine learning algorithms and describe and their limitations.} 
    \item Based on the limitations of the surveyed papers, we identify further research problems to be addressed. {By providing such discussion, we aim to shed light on future directions which utilize machine learning algorithms for FDIAs detection.}
\end{itemize}

The rest of the paper is organized as follows:
Section \ref{sec2} provides the background on power system SE, BDD methods, and FDIAs. Section \ref{sec3} provides details on different FDIA formulations and their impact on power systems. In Section \ref{sec4}, we present traditional defense strategies against FDIAs. We survey different machine learning methods for attack detection in Section \ref{sec5}. Section \ref{sec6} discusses the performance of machine learning algorithms in the context of SE while conclusions are presented in Section \ref{sec7}. Common notations used in the paper are listed in Table \ref{tabpp}.

\begin{table}[t]
\caption{Notations.}
\label{tabpp}
    \centering
    \begin{tabular}{|c|c|}
    \hline
    \textbf{Notation}  
    & \textbf{Parameter}   \\
\hline
$m$  &The number of measurements   \\\hline
$n$  &The number of state variables  \\\hline
$\mathbf{H}$   &$m \times n$ Jacobian matrix representing the topology  \\\hline
$\mathbf{x}$  & $n \times 1$ vector of state variable   \\\hline
$\mathbf{z}$  & $m \times 1$ vector of measurements  \\\hline
$\mathbf{Z}$  & $m \times n$ measurements matrix\\\hline
$\mathbf{e}$  & $m \times 1$ vector of measurement errors, s.t, $\mathbf{z}=\mathbf{H}\mathbf{x}+\mathbf{e}$   \\\hline
$\mathbf{\hat x}$ & $n\times 1$ vector of estimated state variables\\\hline
$\mathbf{W}$ &\makecell{$m\times m$ diagonal matrix, s.t., $w_i,i= \sigma_i^{-2},$ where $\sigma_i^{2}$ \\is the variance of the $i$-th measurement $(1\leq i \leq m)$}\\\hline
$\tau$ & \makecell{Threshold for $L_2$-norm based bad data detection}\\\hline
$\mathbf{z}_a$& $m\times 1$ Measurement vector with bad measurement\\\hline
$\mathbf{a}$ & $m\times 1$ Attack vector, $s.t.,$ $\mathbf{z}_a=\mathbf{z}+\mathbf{a}$\\\hline
$\mathbf{c}$& $n\times 1$ Vector of estimation errors $s.t.,$ $\mathbf{a}=\mathbf{H}\mathbf{c}$\\\hline
$V_i$, $\theta_i$ & Voltage magnitude and phase angle at bus $i$\\ \hline
$g_{ij}, b_{ij}$&\makecell{The real and imaginary parts of the admittance \\of the series branch between bus $i$ and bus $j$}\\
\hline
\end{tabular}
\end{table}

\section{Background}\label{sec2}

\subsection{Power System State Estimation}
{SE enables System Operators (SOs) to optimally manage, plan, and control the power grid. SE is used to assess the system's state, check for anomalous behavior, and indicate if mitigation strategies are necessary to preserve nominal operation. Depending on the power system level that SE is applied, i.e., transmission level or distribution level, different algorithms, assumptions and approximations are employed. The differences between Transmission System (TS) and Distribution System (DS) in terms of SE algorithms are discussed in Sections \ref{s:SET} and \ref{s:SED}. Multiple SE algorithms have been proposed aiming to optimize the computational intensive estimation process and enable its real-time calculation~\cite{soares2019full, 6616007, ghahremani2011dynamic, 336098, 6112697, 5669381, 5871327, haughton2012linear, teng2002using, majumdar2016three, monticelli2000electric, gao2017state}. Despite the plurality of SE methodologies and their application level, the core components of these analyses are fundamentally similar. An outline of the SE process is presented in Algorithm \ref{alg:SE}. 
}
\begin{algorithm}[t!]
\setstretch{1}
\small
\SetAlgoLined
\DontPrintSemicolon
    Input: Parameters, Measurements, Pseudomeasurements, $\tau$ \;
    Output: StateEstimates  \;
     \While{PS.on}{
      Topology = PS.build(Parameters, Measurements) \;
      ObservabilityMatrix = PS.map(Topology, Pseudomeasurements)\;
      \While{error $\geq \tau$}{
        [error, StateEstimates] = PS.stateEstimation(Topology, ObservabilityMatrix)    
        } 
      return StateEstimates \; 
    }
    
\caption{Overview of State Estimation Process.}
\label{alg:SE}
\end{algorithm}

\subsection{Transmission System Modeling for State Estimation} \label{s:SET}

{In this case, it is typically assumed that the system is balanced, overdetermined, i.e., the number of available measurements are more than number of the unknown state variables, and that the system nodes are connected in a mesh topology. These assumptions simplify the  analysis, contrary to DSs which are radially connected, unbalanced, and insufficient measurement points are available (Section \ref{s:SED}). The inputs to the SE are \textit{(i)} the power system parameters (e.g., lines, buses, branches, breaker states, etc.), \textit{(ii)} the collected measurements (e.g., voltages, angles, real/reactive power injections and flows), and \textit{(iii)} pseudomeasurements (e.g., load forecasts, historical data, etc.) which are utilized when insufficient system information is available. }

{The TS model is composed of a set of buses $\mathcal{N} = \{1,\ldots,n\}$, where $n = |\mathcal{N}|$ is the total number of buses. Furthermore, the states of the system at each bus include the voltage magnitude and the phase angle. We denoted the system states using $\mathbf{x} = (x_1,x_2,\ldots,x_n)^T $. Depending on the fidelity of the model, the system measurements can include active and reactive power injections, active and reactive power flows, voltage magnitudes, voltage angles, current magnitudes, etc.~\cite{bandak2014power}. Finally, the set of measurements is denoted as $\mathbf{z}= (z_1,z_2,\ldots,z_m)^T$, where $m$ is the number of measurements.}

{The SE inputs are used in order to build an accurate TS topology and the observability matrix. By inspecting the obervability matrix we can determine which system states are unobservable and derive approximations using the redundancy of the overdetermined system measurements as well as the pseudomeasurements. The calculated results are passed to the main SE routine which iterate until an optimal system solution (based on the imposed constraints) is reached. Solving the SE problem can be a time and resource consuming procedure. Additionally, SE is sensitive to measurement errors which can also impact the algorithm's convergence efficacy. Following, we present the AC SE methodology and demonstrate it as a nonlinear optimization problem. Additionally, in Section \ref{s:DCSE} we present how, in favor of real-time performance and by partially sacrificing the model's accuracy, we derive a linear (DC) model for the SE problem. }

\begin{figure}[t]
\centering{\includegraphics[width=60mm,height= 45mm,keepaspectratio]{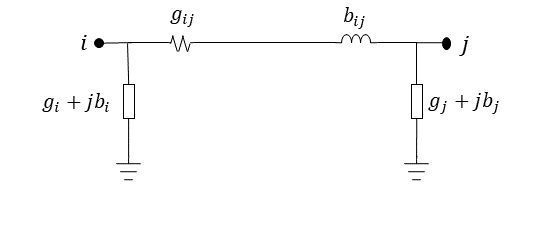}}
\caption{Transmission network element between bus $i$ and bus $j$.}
\label{pic3}
\end{figure}

\subsubsection{AC State Estimation} \label{s:ACSE}

{The AC SE leverages phase angles and voltage magnitudes to construct the system states. Typically, the phase angle at the slack bus is set as the reference, i.e., $\theta_1 = 0$, thus it is not included in the system state vector $\mathbf{x}$. With this assumption, we define the power system states as:}

\begin{equation}
    \mathbf{x} =\left(\theta_{2}, \theta_{3}, \ldots, \theta_{n}, V_{1}, V_{2}, \ldots, V_{n}\right)^T
\end{equation}

{The measurements, $\mathbf{z}$, include bus voltages and angles, as well as, the real and reactive power flows and injections. For each bus $i \in \mathcal{N}$, as depicted in Fig. \ref{pic3}, we have: }
\begin{equation}\label{eq:acpi}
    P_i= V_i \sum_{j = 1}^{n}V_j (g_{ij} cos\theta_{ij}+b_{ij}sin\theta_{ij})
\end{equation}
\begin{equation}\label{eq:qi}
    Q_i=V_i\sum_{j = 1}^{n}V_j (g_{ij} sin\theta_{ij}-b_{ij}cos\theta_{ij})
\end{equation}
{where $P_i$ and $Q_i$ are the real and the reactive power injections at bus $i$, respectively. $g_{ij}$ and $b_{ij}$  are the real and the imaginary part of the nodal admittance matrix element $\mathbf{Y}_{ij}$, and $\theta_{ij} = \theta_i - \theta_j$ is the phase angle difference between buses $i$ and $j$.}

{We utilize a nonlinear function vector $\mathbf{h(x)} = (h_1,h_2,\ldots,h_m)^T$ to represent the relationship as presented in Eq.~(\ref{eq:acpi}) and (\ref{eq:qi}). Thus, we obtain the observation model: $\mathbf{z}= \mathbf{h(x)} + \mathbf{e}$, where $\mathbf{e} = (e_1,e_2,....,e_{m})^T$ is the vector of measurement errors~\cite{monti}. These measurement errors are assumed to be independent to each other and follow the Gaussian distribution $\mathcal{N}(\Vec{0}, \mathbf{W})$, in which $\mathbf{W}$ is the covariance matrix of the measurement errors.}
\begin{equation}
    \mathbf{W}=\operatorname{diag}\left\{\sigma_{1}^{2},\sigma_{2}^{2}, \cdots, \sigma_{2n}^{2}\right\}
\end{equation}

{The Weighted Least Square (WLS) technique is one of the most commonly used methods for SE~\cite{bandak2014power}. In WLS, the estimates are obtained by minimizing the sum of the residual squares as illustrated in Eq. \eqref{eq:sest}: }
\begin{align}
    \begin{split}
        \mathbf{\hat x } &=arg\:\min_{x} \mathbf{J(x)} \\
                &= arg\:\min_{x}(\mathbf{z}-\mathbf{h(x)})^T \mathbf{W}^{-1} (\mathbf{z}-\mathbf{h(x)})
    \end{split}
    \label{eq:sest}
\end{align}

{ The optimization problem presented in Eq. (\ref{eq:sest}) can be solved using the iterative normal equation method~\cite{monticelli2000electric}. At any given point, the solution should satisfy the first order optimal condition of Eq. (\ref{eq:first-order}): }

\begin{equation}\label{eq:first-order}
    \mathbf{G_1(\mathbf{\hat{x}})} = \frac{\partial J(x)}{\partial x}|_{x=\hat{x}} = -\mathbf{{H}}^{T}\left(\mathbf{\hat{x}}\right) {\mathbf{W}^{-1}}\left[{\mathbf{z}} - \mathbf{{h}}\left(\mathbf{\hat{x}}\right)\right]=0
\end{equation}
{where $\mathbf{{H}({x})}=\frac{\partial {h}({x})}{\partial {x}}$ is the Jacobian matrix (Eq. (\ref{Jac})) derived from the function vector $\mathbf{h(x)}$, and $\mathbf{\hat x}$ is the estimated state vector. }

\begin{equation}\label{Jac}
\mathbf{H(x)}=\frac{\partial h(x)}{\partial(x)}=\begin{bmatrix}\frac{\partial h_{1}(x)}{\partial x_{1}} & {\frac{\partial h_{1}(x)}{\partial x_{2}}} & {\cdots} & {\frac{\partial h_{1}(x)}{\partial x_{n}}} \\ {\frac{\partial h_{2}(x)}{\partial x_{1}}} & \frac{\partial h_{2}(x)}{\partial x_{2}} & \cdots & \frac{\partial h_{2}(x)}{\partial x_{n}} \\ \vdots & \vdots & \ddots & {} \\ \frac{\partial h_{m}(x)}{\partial x_{1}} & \frac{\partial h_{m}(x)}{\partial x_{2}} & \cdots & \frac{\partial h_{m}(x)}{\partial x_{n}}
\end{bmatrix}
\end{equation}

\noindent Eq.  (\ref{eq:first-order}) can be iteratively solved using the Newton-Raphson method, and $\mathbf{\hat x}$ can be approximated with:
\begin{align}
   \mathbf{\hat{x}}_{v+1} = \mathbf{x}_{v} + ((\mathbf{G}_2^{T} \mathbf{G}_2)^{-1} \mathbf{G}_2^{T} \mathbf{G}_1)|_{x=x_v}
\end{align}
{where,  $\mathbf{G_2} = \partial^2{J}/\partial{x^2}$ is the Hessian matrix of $\mathbf{J(x)}$ and $v \in \mathcal{N}$ is the iteration step. }

{ Alternative methods, such as the Maximum Likelihood Estimation (MLE) can be employed in order to solve the optimization problem of Eq. (\ref{eq:sest})~\cite{monti}. Additionally, orthogonal methods can be utilized to solve the first optimal condition introduced of Eq. (\ref{eq:first-order})~\cite{monticelli2000electric}. However, the rate and convergence accuracy of these heuristic methodologies rely solely on the system observability matrix characteristics (i.e., rank). }

\subsubsection{DC State Estimation} \label{s:DCSE}

{In order to alleviate the computational burden introduced by nonlinear AC SE, the DC SE model (a linear measurement model) is often considered at the sacrifice of accuracy. The DC SE assumes that the line resistance is negligible compared to the corresponding line reactance, and the phase angle difference between neighboring nodes is small (i.e., zero degrees angle difference). Also, the voltage magnitudes are assumed to be $1$ \textit{p.u.}. Thus, dissimilar to the AC SE, in DC SE the system states are composed only from the phase angles $x =\left(\theta_{2}, \theta_{3}, \ldots, \theta_{n}\right)^T$. Moreover, since the reactive power flow between buses is negligible and the reactive power injections at every bus depend on the line susceptance, only active power flows and injections are utilized in DC SE: }
\begin{equation}\label{eq:pi}
    P_i= \sum_{j = 1, j\not = i}^{n} b_{ij}(\theta_{i} - \theta{j})
\end{equation}
{ Thus, the observation model in DC SE can be formalized as: }
\begin{equation}\label{eq:dcse}
   \mathbf{z = Hx+e}
\end{equation}
{ where, $\mathbf{H_{ii}} = \sum_{j=1, j\not=i}^{j=n} {b_{ij}}$ and $\mathbf{H_{ij}} = -b_{ij}$. $\mathbf{z} = (P_1,\cdots,P_n)^T$, and $\mathbf{e}=(e_1,\cdots,e_n)^T$ follows the same assumptions as in the AC SE. Leveraging WLS to solve Eq.  (\ref{eq:dcse}), we obtain the following objective function formulation: }
\begin{align}\label{eq:dcsest}
    \begin{split}
        \mathbf{\hat x} &=arg\:\min_{x} J(\mathbf{x}) \\
                &= arg\:\min_{x}(\mathbf{z-Hx})^T \mathbf{W}^{-1} (\mathbf{z-Hx}) 
    \end{split}
\end{align}
{The solution satisfies the following requirement: }
\begin{equation}\label{eq:dcfirst-order}
    G_1(\mathbf{\hat{x}}) = \frac{\partial J(x)}{\partial x}|_{x=\hat{x}} 
        = -\mathbf{{H}}^{T} {\mathbf{W}^{-1}}\left[\mathbf{{z} - H\hat{x}}\right]=0
\end{equation}
\noindent {which can be simplified to, }
\begin{equation}
    \mathbf{\hat{x}} = (\mathbf{H}^{T}\mathbf{WH})^{-1}\mathbf{H}^{T}\mathbf{Wz}
\end{equation}

\subsubsection{Dynamic State Estimation} \label{s:DynSE}
{SOs have extensively used both AC and DC static SE in order to monitor TS operation and manage energy generation. However, these SE algorithms rely on bus voltage and angle measurements as well as active and reactive power injections to calculate the system state estimates. The disadvantage of these methods is that the system state approximation -- using either nonlinear or linear models -- depends on low-update rate steady-state measurements e.g., Supervisory Control and Data Acquisition (SCADA)~\cite{6616007}. The current transmission infrastructure advancements, with the integration of wind and solar generation, require improved estimation algorithms able to capture the dynamic system behavior~\cite{ghahremani2011dynamic}. In order to address the aforementioned AC and DC SE pitfalls, and account for the dynamic and intermittent nature of TS with renewable penetrations, Dynamic SE (DSE) algorithms have been proposed.}

{ The  initial implementations of DSE algorithms although they could acutely reflect the system's transient behavior, they still suffered from the disadvantages of the traditional SE methodologies~\cite{4840192, 4745312, 6112697}. For example, to achieve faster convergence rates, nonlinear models had to be linearized causing significant approximation errors and Jacobian system matrices had to be recalculated at every iteration step yielding excessive computational overheads~\cite{6112697}. In order to overcome linearization errors and computational intensive matrix operations, recent works have opted for improved SE methodologies leveraging  Kalman filtering techniques. In~\cite{5669381}, an UKF method for DSE is introduced, which overcomes the aforementioned drawbacks and avoids the high order derivative calculations in favor of real-time performance. Other works have incorporated high data-rate sampling measurements from PMUs to increase the robustness of their estimations ~\cite{5871327}. For instance, the authors in~\cite{6616007}  showed improvements in estimation accuracy, algorithm convergence, and minimized the estimation complexity. Their algorithm allows leveraging UKF, PMU measurements, as well as a decentralized SE approach, demonstrating a practical implementation for TS DSE. }

\subsection{Distribution System Modeling for State Estimation} \label{s:SED}

{In the past, SE was applied exclusively on the transmission level since the distribution level can be simplified to a lumped passive load structure. However, with the deployment of distributed generation, distributed energy resources (DERs), microgrids, electric vehicles, and energy storage systems, the development of comprehensive algorithms that account for bidirectional power flow between transmission and distribution levels is imperative. The first Distribution System State Estimators (DSSE) are adaptations of the corresponding TS counterparts ~\cite{336098}. However, DS architectures differ significantly from transmission networks. }

{DSs are radially connected and their interconnections typically present high resistance to reactance (R/X) ratios. On the other hand, TSs are connected in lattice-based formations, aiding redundancy, and their line resistances are  negligible. Second, there are fewer measurement points in DSs when compared to TSs, and even when measurements are available, they are usually collected deficiently (every $15$ minutes or even longer). Also, measurements might not be time-synchronized and can be inaccurate (due to improper connections or not calibrated meters). Thus, relying on pseudomeasurements for DSSE is a common practise. Furthermore, DSs are constantly changing by  integrating distributed generation units, loads, and prosumers, thus DSSE algorithms should be able to account for such characteristics. Finally, TS are treated as perfectly \textit{balanced systems} by SE algorithms; such algorithms cannot be applied for DS topologies since they present serious imbalances between phases and require three-phase modeling. }

\subsubsection{State Estimation for Unbalanced System Operation}

{The dynamic behavior of DSs, furnishing variable power penetrations and demands at every system bus, generates load flow differences between phases. Thus, for practical DSs solving the unbalanced three-phase problem is required to perform DSSE. For instance, in~\cite{gao2017state} the authors solve the DSSE problem utilizing unbalanced single-phase and two phase measurement models. Eqs. (\ref{eq:3P}) and (\ref{eq:3Q}) demonstrate the active and reactive power flows for the three-phase system model at bus $i$ and phase $p$. In the aforementioned equations, $V$ is the voltage and $\theta_{i k}^{p m}$ present the phase angle difference between bus $i$ with phase $p$ and bus $k$ with phase $m$. $g_{i k}^{p m}$ and $b_{i k}^{p m}$ are the corresponding real and imaginary parts of the admittance matrix representing the conductance and susceptance for each bus, respectively. }

\begin{equation}
\label{eq:3P}
P_{i}^{p}\!\!=\!\!\left|V_{i}^{p}\right| \sum_{k=1}^{N} \sum_{m \in\{a, b, c\}}\!\!\!\!\!\!\left|V_{k}^{m}\right|\left(g_{i k}^{p m} \cos\!\left(\theta_{i k}^{p m}\right)\!+\!b_{i k}^{p m} \sin\! \left(\theta_{i k}^{p m}\right)\right)
\end{equation}

\begin{equation}\label{eq:3Q}
Q_{i}^{p}\!\!=\!\!\left|V_{i}^{p}\!\right| \sum_{k=1}^{N} \!\sum_{m \in\{a, b, c\}}\!\!\!\!\!\!\!\!\left|V_{k}^{m}\right|\left(g_{i k}^{p m} \sin \!\left(\theta_{i k}^{p m}\right)\! -\! b_{i k}^{p m} \cos\! \left(\theta_{i k}^{p m}\right)\right)\end{equation}

{Additionally, the line-to-line voltage as well as the real power injection of the two-phase measurement model are demonstrated in Eqs. (\ref{eq:2v}) and (\ref{eq:2p}). To calculate the mentioned bus voltages and power injections,  \textit{(i)} the three-phase power injection, the phase-to-neutral voltage magnitude, and the magnitude of current injection at the current substation, in addition to, \textit{(ii)} the two-phase voltage magnitudes and power injections at every distribution center-tapped transformer are necessary.}

\begin{equation}
\begin{split}
\label{eq:2v}
\left|V_{i}^{p m}\right|_{\text {meas}}\!\!\!\!\!&=\sqrt{\left|V_{i}^{p}\right|^{2}+\left|V_{i}^{m}\right|^{2}-2\left|V_{i}^{p}\right|\left|V_{i}^{m}\right| \cos \left(\theta_{i}^{p}-\theta_{i}^{m}\right)}\\
&+e_{\left|V_{i}^{p m}\right|}
\end{split}
\end{equation}

\begin{equation}\label{eq:2p}
\begin{aligned}
P_{i \text { meas }}^{p m} \!\!\!&=\!\!\left|V_{i}^{p}\right| \sum_{k=1}^{N} \sum_{n \in\{a, b, c\}}\!\!\!\!\!\!\!\left|V_{k}^{n}\right|\!\left(g_{i k}^{p n} \cos\!\left(\theta_{i k}^{p n}\right)\!\!+\!\!b_{i k}^{p n} \sin\!\left(\theta_{i k}^{p n}\right)\right) \\
&-\!\left|V_{i}^{m}\right|\!\!\!\!\!\!\!\!\!\! \sum_{k=1 n \in\{a, b, c\}}^{N}\!\!\!\!\!\!\!\!\!\!\!\left|V_{k}^{n}\right|\!\left(g_{i k}^{p n} \cos\! \left(\theta_{i k}^{m n}\right)\!\!+\!\!b_{i k}^{p n} \sin\! \left(\theta_{i k}^{m n}\right)\right)\!\!\\
&+\!e_{P_{i}^{p m}}
\end{aligned}
\end{equation}

{Furthermore, when the phase-to-neutral voltage magnitudes and the real power injection measurements are available -- assuming ideal center-tapped and single-phase transformers (i.e., the transformer losses are negligible) -- we can acquire the following single-phase measurement equations.}

\begin{equation}\left|V_{i}^{p}\right|_{m e a s}=\left|V_{i}^{p}\right|+e_{\left|V_{i}^{p}\right|}\end{equation}

\begin{equation}P_{i \text { meas }}^{p}=P_{i}^{p}+e_{P_{i}^{p}}\end{equation}

\begin{equation}Q_{1 \text { meas }}^{p}=Q_{1}^{p}+e_{Q_{1}^{p}}\end{equation}

{Performing the Kron reduction method on the initial four-wire matrix which also includes the line-to-neutral impedances, the simplified (row and column reduced) line impedance  can be obtained by using the resistance ($R$) and reactance ($X$) of the line. A three-phase (a, b, and c) line impedance matrix between bus $i$ and $j$ can be calculated by utilizing Eq. (\ref{eq:unbalance1}).}

\begin{equation}\label{eq:unbalance1}
\begin{split}
\mathbf{Z}_{Imp(abc,ij)} &= \mathbf{R}_{abc,ij}+ j\mathbf{X}_{abc,ij}\\
&= \begin{bmatrix}
 &  \mathbf{Z}_{Imp(aa,ij)}^{n},\mathbf{Z}_{Imp(ab,ij)}^{n},\mathbf{Z}_{Imp(ac,ij)}^{n} & \\ 
 & \mathbf{Z}_{Imp(ba,ij)}^{n},\mathbf{Z}_{Imp(bb,ij)}^{n},\mathbf{Z}_{Imp(bc,ij)}^{n} & \\ 
 &  \mathbf{Z}_{Imp(ca,ij)}^{n},\mathbf{Z}_{Imp(cb,ij)}^{n},\mathbf{Z}_{Imp(cc,ij)}^{n} & 
\end{bmatrix} 
\end{split}
\end{equation}

{This methodology can be applied irrespective of the system modeling being single-phase, two-phase, or three-phase. For example, if we opt for a single-phase model, the corresponding row and column of the other two phase will be zero.}

{Furthermore, the branch voltages and branch currents modeling is shown in Eqs. \eqref{eq:voltageabc} and \eqref{eq:currentabc}, respectively. The mentioned branch voltage and current modeling allows for direct use in voltage-based or branch current-based SE methods~\cite{teng2002using}.  }

\begin{equation}
   \mathbf{V_{abc,ij}} = \begin{bmatrix}
V_{ai}\\ 
V_{bi}\\ 
V_{ci}
\end{bmatrix} -  \begin{bmatrix}
V_{aj}\\ 
V_{bj}\\ 
V_{cj}
\end{bmatrix} 
\label{eq:voltageabc}
\end{equation}

\begin{equation}
    \mathbf{I_{abc,ij}} = \begin{bmatrix}
I_{a,ij}\\ 
I_{b,ij}\\ 
I_{c,ij}
\end{bmatrix}
\label{eq:currentabc}
\end{equation}

{Other methods leverage WLS to construct a linear SE model for unbalanced three-phase systems~\cite{haughton2012linear}. For this linear approximation the bus voltages and branch currents as well as the active and reactive power flow measurements are essential for the three-phase unbalanced system model. Further, the SE algorithm requires timely synchronized phasor measurements. We demonstrate the measurement vector, state vector, and the process matrix $\mathbf{H}$ in Eqs. (\ref{eq:Zequation}), (\ref{eq:Xequation}), and (\ref{eq:Hequation}), where the subscripts $r$ and $i$ are the real and imaginary values. The system residuals for unbalanced operation are formulated in Eq. (\ref{eq:Requation}).} {All the aforementioned differences in system modelling make DSSE an arduous and computational intensive process limiting its real-time applicability. }

\begin{equation}
\label{eq:Zequation}
   \mathbf{z}=\mathbf{z}_{r}+ j \mathbf{z}_{i}
\end{equation}
\begin{equation}
\label{eq:Xequation}
   \mathbf{x}=\mathbf{x}_{r}+ j \mathbf{x}_{j}
\end{equation}
\begin{equation}
\label{eq:Hequation}
    \mathbf{H}=\mathbf{H}_{r}+ j \mathbf{H}_{j}
\end{equation}

\begin{equation}
\label{eq:Requation}
\mathbf{r}_{r}+j \mathbf{r}_{i} = \mathbf{z}_{r}+ j \mathbf{z}_{i} - (\mathbf{H}_{r}+j \mathbf{H}_{i}) (\mathbf{x}_{r}+j \mathbf{x}_{i} )
\end{equation}

\subsection{Bad Data Detection and Identification}\label{sec2.2}
{With bad data injected during the SE, the states might not be accurate, which could lead to wrong decision making and economic losses.} Therefore, it is necessary to sanitize the measurements by removing the bad data. Some bad data such as negative voltage magnitudes can be easily removed before the SE process. However, other require sophisticated methods to detect, identify, and remove them from the true measurement vectors.

\subsubsection{Bad Data Detection}
{The goal of BDD is to determine whether bad data exist in the measurement vectors \cite{merrill1971bad, handschin1975bad}. The Chi-Square test is a statistical method widely used for this process. Chi-square assumes that the distribution of measurements follows a Gaussian distribution. Thus, the test statistic $\mathbf{J(\hat{x})}$ (calculated in Eq. (\ref{sg1})) follows Chi-square distribution when there exist no bad data~\cite{zhao2016enhanced}}:
\begin{equation}\label{sg1}
   \mathbf{ J(\hat{x})}=\sum_{i=1}^{m} \frac{\left(\mathbf{z}_{i}-\mathbf{h}_{i}(\mathbf{\hat{x}})\right)^{2}}{\sigma_{i}^{2}}
\end{equation}
where $\mathbf{r}_i= \mathbf{z}_i - \mathbf{h}_i(\mathbf{\hat x)}$ is known as the residual. If $\mathbf{J(\hat{x})}$ is larger than a predetermined threshold, then bad data exist in the measurements.

\subsubsection{Bad Data Identification} \label{s:BDD_ident}
{The goal of the bad data identification procedure is to determine which set of measurements contains bad data \cite{lin2018highly, aghamolki2018socp}. The Largest Normalized Residual (LNR) is one of the most commonly used methods for bad data identification~\cite{van1985bad}.} Similar to the Chi-square test method, LNR assumes that the bad measurements have large residuals. The following steps detail the process of identifying bad data using LNR: 
\begin{enumerate}[leftmargin=3em,labelsep=1em]
   \item   Calculate the gain matrix $\mathbf{G}$ and the covariance matrix $\mathbf{\Omega}$:
    \begin{equation}
       \mathbf{ G}=\mathbf{H}^T\cdot \mathbf{W}\cdot \mathbf{H}
    \end{equation}
    \begin{equation}
        \mathbf{\Omega (\hat x)} = \mathbf{W}-\mathbf{H(\hat x)}\cdot \mathbf{G}^{-1}\cdot \mathbf{H}^T\mathbf{(\hat x)}
    \end{equation}
    \item Calculate the normalized residuals after solving the estimation problem using the WLS method:
    \begin{equation}
    \mathbf{r}_i^n= \frac{|\mathbf{r}_i|}{\sqrt{\mathbf{\Omega_{i}}}} \:\; i=1,2..m
    \end{equation}
    \item Find the maximum value $r_{max}^n$ of $r_i^n$ for $i= 1,2, \dots ,m$.
    \item Compare the LNR with a pre-determined threshold $\tau$. If 
    $\parallel \mathbf{r}_{max}^n \parallel > \tau $, the corresponding measurement is assumed to be bad (modified).
    \item Remove the suspected bad measurement from the measurement set and go to step one. 
\end{enumerate}

Although residual-based methods are widely used, it has been demonstrated that they cannot efficiently detect FDIAs~\cite{dan2010stealth}.

\subsection{False Data Injection Attacks in Power Systems}

{FDIAs are a class of cyberattacks which can bypass BDD mechanisms, and aim to compromises the data integrity of power system measurements. SOs utilize SE on both the transmission and the distribution level. SE results serve as inputs to other crucial power system services (e.g., optimal power flow, economic dispatch, demand-response, contingency analysis, etc.), thus their validity is of paramount importance. Ensuring accurate results requires meticulous line interconnection and topology modeling as well as scaleable and dynamic algorithms. Furthermore, efficient SE algorithms which can harness pseudomeasurements (based on historical data or forecasts) and comply with the real-time system operational requirements are crucial. Attackers, either intrusively (e.g., having physical access to a grid asset which reports measurements) or non-intrusively (e.g., by spoofing a communication channel over which power system measurements are propagated) can maliciously modify and inject false data in the system. Typically, a FDIA is formulated as follows: }
\begin{equation}
    \mathbf{z_a}= \mathbf{z} + \mathbf{a} 
\end{equation}
{where $\mathbf{z_a}$ is the tampered measurement vector, $\mathbf{z}$ is the true measurement vector, and $\mathbf{a}$ is a non-zero attack vector added to the true measurements.} 

{In order to bypass BDD (i.e., not affect residuals), the attack vector $\mathbf{a}$ is constructed as a linear combination of the column vectors of the Jacobian $\mathbf{H}$ matrix, that is, $\mathbf{a} = \mathbf{H}\mathbf{c}$, where $\mathbf{c}$ is an arbitrary $n\times 1$ non-zero vector. The attack vector is constructed as follows:}
\begin{equation}
    \begin{bmatrix}
    {a_{1}} \\ {a_{2}} \\ {\cdot} \\ {a_{m}}\end{bmatrix}_{m\times1}= c_{1} \begin{bmatrix}{h_{11}} \\ {h_{21}} \\ {\cdot} \\ {h_{m 1}}\end{bmatrix} +\cdots+c_{n}  \begin{bmatrix}{h_{1 n}} \\ {h_{2 n}} \\ {\cdot} \\ {h_{m n}}\end{bmatrix}
\end{equation}
\begin{equation}\label{z_a}
    \mathbf{z_a}=\mathbf{H}(\mathbf{x}+\mathbf{c})
\end{equation}
and the new estimated state $\mathbf{\hat{x}_{a}}$ is equal to:
\begin{equation}
    \mathbf{\hat{x}_{a}}= \mathbf{\hat{x}}+\mathbf{c}
\end{equation}
The value of $\mathbf{c}$ should not exceed the maximum alterable tolerance of any measurement to avoid triggering alarms and draw the the grid operator's attention~\cite{liang2017false}. { Following this procedure, $\mathbf{z_a}$ produces the same residual as the real measurement vector $\mathbf{z}$, and thus bypasses the residual-based BDD (for the DC SE model). }
\begin{equation}\label{ra1}
\begin{split}
\mathbf{r_{a}}&= \parallel \mathbf{z_{a}}- \mathbf{H}\mathbf{\hat{x}_{a}} \parallel \\
& = \parallel \mathbf{z}+\mathbf{a} - \mathbf{H}(\mathbf{\hat{x}}+\mathbf{c})\parallel\\
& = \parallel \mathbf{z}+\mathbf{a}-\mathbf{H}\mathbf{\hat{x}}-\mathbf{H}\mathbf{c}\parallel \\
& = \parallel \mathbf{z}-\mathbf{H}\mathbf{\hat{x}}+(\mathbf{a}-\mathbf{H}\mathbf{c})\parallel\\
& =\parallel \mathbf{z}-\mathbf{H}\mathbf{\hat x}\parallel =\mathbf{r} 
\end{split}
\end{equation}

{In Eq. (\ref{ra1}) we prove that if $\mathbf{a}=\mathbf{H}\mathbf{c}$, then $\mathbf{r_{a}}= \mathbf{r}$, indicating that the attack succeeds without changing the measurement residual or triggering the BDD.
FDIA formulation for the AC SE is similar to the DC SE case. The attack bypasses the BDD if $\mathbf{a}= \mathbf{h(\hat x_{a})}-\mathbf{h(\hat x)}$, and therefore the residual remains unaltered (Eq.~\eqref{11}):}
\begin{equation}\label{11}
    \begin{split}
       \mathbf{r_{a}} &= \parallel \mathbf{z_a} - \mathbf{h}(\mathbf{\hat x_{a}})\parallel\\
        &=\parallel \mathbf{z}+\mathbf{a}-\mathbf{h}(\mathbf{\hat x_{a}})+\mathbf{h}(\mathbf{\hat x)} -\mathbf{h}(\mathbf{\hat x})\parallel\\
       &= \parallel \mathbf{z}-\mathbf{h}(\mathbf{\hat x})+\mathbf{a}-\mathbf{h}(\mathbf{\hat x_{a}})+\mathbf{h}(\mathbf{\hat x})\parallel= r
    \end{split}
\end{equation}

{
Many researchers provide use cases where the TS SE can be maliciously manipulated if PMU data, Remote Terminal Units (RTUs) data or SCADA measurements are compromised, as well as how the corresponding FDIAs can be constructed~\cite{liu2011false, teixeira2011cyber, hug2012vulnerability, deng2016false}. Due to the differences between TS and DS modeling and operation, the SE mechanisms can differ significantly as discussed in Sections \ref{s:SET} and \ref{s:SED}. The heavily interconnected  DS topology, the number of insufficient measurement points, dynamic and unbalanced DS operation complicate the DSSE process. Attackers can leverage the elaborate DSSE to mount FDIAs and avoid detection. Research works discussing FDIAs which target DS have been reported~\cite{choeum2019oltc, deng2018false, zhang2020false}. Detecting and mitigating FDIAs is a field of ongoing research. An overview of the state-of-the-art methodologies leveraging machine learning is discussed in Section \ref{sec5}. }

\section{False Data Injection Attack Settings and Impacts}\label{sec3}
{In this section, we provide a brief overview of how FDIAs can be launched according to the attack knowledge settings (summarized in Table~\ref{tabl}) and discuss their potential impacts on power systems (summarized in Table \ref{tab15}).} 

\subsection{FDIAs Settings}
Typically, system knowledge includes meter measurement data, the Jacobian matrix or system topology, system parameters, and control commands (e.g., switch states). Moreover, to compromise the DS SE the attacker should know the state estimates to successfully launch a FDIA. Based on the attacker's knowledge, attackers can be classified into two categories: \textit{(i)} attackers with full system knowledge, and \textit{(ii)} attackers with incomplete or partial system knowledge. Full system knowledge enables the attacker to design FDIAs that will not trigger detection mechanisms. On the other hand, in the case of incomplete or partial information, the attackers may not know the exact system topology (e.g., Jacobian matrix). Thus, attackers first need to approximate this crucial information (i.e., topology matrix) leveraging meter measurements or historical data, before a stealthy FDIA can be launched. 

\begin{table}[t]
\caption{FDIAs Categories.}
\label{tabl}
 \centering
    \begin{tabular}{|c|c|c|}
    \hline
\textbf{FDIAs Categories}  & \textbf{References} & \textbf{Examples}  \\
\hline
Attack with full knowledge &  ~\cite{liu2011false, boba, sou, mallat1993matching}& \makecell {Access specific\\ meters, minimize \\the number \\of attacked meters}  \\
\hline
\makecell{Attack with incomplete \\knowledge}  & ~\cite{18, kekatos2014grid, kim2015subspace, yu2015blind, 20, liu2015modeling, deng2018false}& \makecell {Use online and\\ offline data,\\ utilize market\\ price data}\\
\hline
\end{tabular}
\end{table}
\subsubsection{FDIAs with Full System Knowledge}
The concept of FDIAs in the power grid, originally introduced by Liu \textit{et al}.~\cite{liu2011false}, investigated two different FDIAs scenarios: \textit{(i)} attacks with limited access to meters, and \textit{(ii)} attacks with limited resources to compromise a large number of meters. In the first scenario, the attacker could only compromise $k$ specific meters due to different security requirements of each meter. For the attack to have considerable impact, the authors assume $k\geq m-n+1$, where $m$ is the number of measurements and $n$ is the number of states. In the second scenario, the authors assume that there are no protected meters, but the attacker has limited resources and could only compromise a limited number of meters. Due to resource constraints, the attacker could not compromise more than $k$ meters. In both scenarios, the authors prove that the attacker could systematically and efficiently construct attack vectors which can modify the SE results without being detected. Both scenarios are experimentally demonstrated on IEEE $9, 27,$ and $300$ bus test cases. The simulation results illustrate the significant impact of FDIAs (e.g., blackouts in large geographic areas).  

Sou \textit{et al}. study how the minimum set of meters -- required to compromise the system -- can be found~\cite{sou2011electric}. The authors assume that there are no empty measurements, i.e., all the rows of the observation model matrix $\mathbf{H}$ are non-zero. In their work, the attacker intends to spoof a specific measurement, e.g., the $k$-th measurement. To avoid detection, the attacker also modifies other measurements according to Eq. (\ref{z_a}). Thus, the attacker's objective is to minimize the number of compromised meters in order to reduce the attack cost and detection risk. The sparsest stealthy FDIAs problem formulation is the following:

\begin{equation}\label{75}
\begin{array}{l}
\alpha_{k}=\min :\|\mathbf{H c}\|_{0} \\
\text { Subject to }: \mathbf{H}(k,:) \mathbf{c}=1
\end{array}\end{equation}
where $\alpha_k$ is defined as the security index of the $k$-th measurement, i.e., the minimum number of measurements required to be compromised for a stealthy FDIA to spoof the $k$-th measurement. Multiplying by a constant $\mathbf{c}$, the attacker can tamper the $k$-th measurement with any value. The security index of the measurements helps the SO to understand the data manipulation patterns and allocate protective resources effectively. {In order to solve the optimization problem of Eq. ({\ref{75}}), various methods have been proposed, such as the Mixed-Integer Linear Programming (MILP) method and the matching pursuit methodology~\cite{sou2011electric, mallat1993matching}.}

\subsubsection{FDIAs with Partial or Incomplete Knowledge}

In~\cite{18}, the authors study FDIAs with incomplete transmission line admittance information, i.e., the attacker does not possess an accurate version of the matrix $\mathbf{H}$. As a result, the attacker does not know the exact values of the transmission line admittance for any part of the power grid topology. However, the attacker could build probability distributions and infer the unknown line admittance with offline and online information. The offline information relies on historical measurements, while the online information is collected by deploying meters or PMUs in the system. The authors compare the impact and detection probability of such attacks against full knowledge FDIAs. The simulation results demonstrate that the attacker could still launch successful FDIAs even with incomplete system information.

Other researchers investigate data-driven approaches to build the Jacobian matrix $\mathbf{H}$ and launch FDIAs, referred to as blind FDIAs~\cite{ kim2015subspace, yu2015blind}. In blind FDIAs, no additional knowledge (except system measurements) is required, and the attack is performed utilizing the equivalent $\mathbf{H}$ matrix constructed in accordance to the acquired measurements. The measurements can be obtained either by direct access to the system or by spoofing the system for a short time period.

Kim \textit{et al}. apply Singular Value Decomposition (SVD) to exploit the subspace of matrix $\mathbf{Z}$ and construct the grid topology~\cite{kim2015subspace}. $\mathbf{Z}$ is constructed using a sample of the system measurements over a period $t$ where the $i_{th}$ row represents the measurements at time $i$:
\begin{equation}\label{eqn:zmeasure}
\mathbf{Z}=\begin{bmatrix}z_{11}&z_{12}&\cdots &z_{1m} \\z_{21}&z_{22}&\cdots &z_{2m} \\\vdots & \vdots & \ddots & \vdots\\z_{t1}&z_{t2}&\cdots &z_{tm}\end{bmatrix}
\end{equation}
The covariance matrix of $\mathbf{Z}$, $\mathbf{\Sigma}_Z$, is computed as follows: 
\begin{equation}
    \mathbf{\Sigma}_{\mathbf{Z}} \triangleq {E}[({\mathbf{Z}}-{E}[{\mathbf{Z}}])({\mathbf{Z}}-{E}[{\mathbf{Z}}])^{T}]= {\mathbf{H}} \mathbf{\Sigma}_{\mathbf{{x}}} {\mathbf{H}}^{T}+{\sigma^{2}\mathbf {I}}
\end{equation}
where, $\sigma^{2} \mathbf{{I}}$ is the covariance matrix of the error vector $e$ ($\mathbf{{z}={H}{x}+{e}}$) and $\mathbf{\Sigma}_{\mathbf{x}}$ is the covariance matrix of the state vector $\mathbf{x}$. The basis matrix of $\mathbf{{H}} \mathbf{\Sigma}_{\mathbf{{x}}} \mathbf{{H}}^{T}$ is calculated by applying SVD to $\mathbf{\Sigma}_Z$, i.e., by finding a unitary matrix $\mathbf{U}$, a rectangular diagonal matrix $\mathbf{\Lambda}$, and a unitary matrix $\mathbf{V}$ such that $\mathbf{\Sigma}_{\mathbf{Z}} = \mathbf{U} \Lambda \mathbf{V}^T$. The $n$ columns of the unitary matrix $\mathbf{U}$ are equivalent to the eigenvectors of matrix $\mathbf{{H}} \mathbf{\Sigma}_{\mathbf{{x}}} \mathbf{{H}}^{T}$ which form the basis of the column space of $\mathbf{{H}} \mathbf{\Sigma}_{\mathbf{{x}}} \mathbf{{H}}^{T}$. Since the column space of $\mathbf{{H}} \mathbf{\Sigma}_{\mathbf{{x}}} \mathbf{{H}}^{T}$ is equivalent to the column space of $\mathbf{H}$, the $n$ columns of $\mathbf{U}$ also form a basis of the column space of $\mathbf{H}$. Thus, the attacker can construct a potential attack vector $\mathbf{a}$ using matrix $\mathbf{U}$. 

Similarly, Yu \textit{et al}. leverages Principal Component Analysis (PCA) to construct blind FDIAs ~\cite{yu2015blind}. PCA is a dimensionality reduction and data transformation method used to reduce a large set of variables to a small set while retaining the critical information of the original set. The authors apply PCA to $\mathbf{z}$, which is the measurement vector, and obtain a transformation matrix $\mathbf{H_{pca}}$, as well as, the principal components vector $\mathbf{\Tilde{x}}$, illustrated in Eq. (\eqref{fr}:

\begin{equation}\label{fr}
\begin{aligned}
    \mathbf{z} & \approx\left[\begin{array}{cccc}
    \mathbf{\Tilde{p}_{1}} & \mathbf{\Tilde{p}_{2}} & \cdots & \mathbf{\Tilde{p}_{n}} \\
    \end{array}\right]\left[\begin{array}{c}
    \mathbf{\Tilde{x}_{1}} \\
    \vdots \\
    \mathbf{\Tilde{x}_{n}} 
    \end{array}\right] 
    \equiv & \mathbf{H_{pca}} \mathbf{\Tilde{x}_{pca}}
    \end{aligned}\end{equation}

\noindent where, $\mathbf{z}$ is a $m \times 1$ overdermined measurement vector, $\mathbf{H}_{pca}$ is an $m \times n$ matrix with $n$ eigenvectors ($\mathbf{\Tilde{p}_{i}}$), and $\mathbf{\Tilde{x}}_{pca}$ is the $n \times 1$ principal components vector. 
The PCA reduced $\mathbf{H}_{pca}$, is leveraged for the construction of the blind FDIA and the formation of the attack measurement vector $\mathbf{ {z}_{a}}$ as described in Eq.~\eqref{eqn:apca} and Eq.~\eqref{eqn:zpca}.

\begin{equation}\label{eqn:apca}
\mathbf{a_{pca}}=\mathbf{H_{pca}}\times \mathbf{c}
\end{equation}
\begin{equation}\label{eqn:zpca}
      \mathbf{ {z}_{a}}=\mathbf{{z}}+\mathbf{{a}_{pca}}
\end{equation}

Teixeira \textit{et al}. study stealthy FDIAs in dynamic systems where the $\mathbf{H}$ matrix is changing ~\cite{teixeira2011cyber}. The attack is constructed following ~\cite{sou2011electric}. The authors mathematically prove the possibility of local stealthy FDIAs. If the changes in $\mathbf{H}$ do not affect the compromised measurements, the attack vector -- constructed using the original $\mathbf{H}$ -- remains stealthy after any system change. Furthermore, the authors empirically study the impact that the magnitude of an attack vector can introduce on the success rate of the attack. They conduct experiments utilizing the IEEE $39$ bus test case with Energy Management System (EMS) software including SE and residual-based BDD. The results validate that even large attack vectors can bypass the detection.

Kekatos \textit{et al}. propose an algorithm leveraging Locational Marginal Prices (LMPs) which is computed from a network-constrained economic dispatch problem to recover the grid Laplacian with a regularized MLE ~\cite{kekatos2014grid, anubi202enhanced, anubi2019resilient}. Esmalifalak \textit{et al}. propose an Independent  Component Analysis (ICA) algorithm to estimate the $\mathbf{H}$ matrix and system topology by observing the power flow measurements ~\cite{esmalifalak2011stealth}. Similarly, Liu \textit{et al}.  show that the attacker could launch an attack in a local region possessing only local system information ~\cite{liu2015modeling}.

{Deng \textit{et al}. propose a practical FDIA model against SE in DSs, where the attacker can successfully launch FDIAs with partial system information \cite{deng2018false}. The authors illustrate how the attacker could estimate the system states based on a small amount of power flow or power injection measurements. The proposed method reduces the cost of obtaining system states, making FDIAs more realistic against SE on the distribution level. The proposed model is demonstrated on an IEEE test feeder. The results show that attacks can effectively compromise the SE avoiding detection.}

\subsection{Impacts of FDIAs on Power Systems}
\begin{table}[t]
\caption{Impacts of FDIAs on power system.}
\label{tab15}
\centering
\begin{tabular}{|c|c|c|}
\hline
\textbf{Impacts}  & \textbf{References} & \textbf{Examples}  \\
\hline
\makecell{The power
grid operation} &  ~\cite{yuan2011modeling, review, rahman2014impact}&\makecell{Load\\ redistribution \\attack}\\
\hline
\makecell{Distributed energy\\ routing
process}  & ~\cite{lin2012false, teixeira2011cyber}& \makecell{Energy \\deceiving attack}\\
\hline
\makecell{Affect the operation of the\\
deregulated electricity market} & ~\cite{xie2010false, xie2011integrity} & \makecell{Economic \\ attack }\\
\hline 
\end{tabular}
\end{table}
FDIAs can cause significant economic or physical impacts on the power system. In this section, we review the effects of FDIAs and summarize them in Table \ref{tab15}.

\subsubsection{Load Redistribution Attack}
Yuan \textit{et al}. propose a particular type of FDIAs, called Load Redistribution (LR) attack, which targets the Security-Constrained Economic Dispatch (SCED) and can potentially affect the power grid operation ~\cite{yuan2011modeling}. The power system uses SCED to reduce the total system operation cost by properly re-dispatching the generation output. Due to LR attacks, the SCED provides incorrect solutions based on corrupted state estimates and drives the system to infeasible operating states. Moreover, the LR attacks can potentially cause load shedding events  immobilizing any immediate corrective action~\cite{yuan2011modeling, rahman2014impact}.

\subsubsection{Energy Deceiving}
Liu \textit{et al}. studies a new variation of FDIAs named energy deceiving attacks which target the routing process of energy distribution~\cite{lin2012false}. The authors introduce a distributed energy routing scheme to find the optimal route for energy flow between nodes of the grid. Each node could be either an energy consumer or an energy producer. To distinguish different nodes, a measurement tool is used (e.g., a smart meter). All nodes communicate with each other to share information such as measurements, requests, and demands. The energy deceiving attack is conducted by spoofing the information exchanged between nodes. Malicious energy information or malicious link-state information is injected into the energy request and response messages of the nodes. A successful attack can manipulate the memory of a measurement tool and inject the false demand and supply messages to the grid. The authors analyze the impact of the energy deceiving attack based on the proposed method and conclude that the attack would create imbalances between demand and supply. As a result, the cost of energy distribution can severely increase. 

\subsubsection{Economic Attack}

In terms of impacts on economic operations, Xie \textit{et al}. demonstrate how FDIAs affect the energy market ~\cite{xie2010false}. {Real-time market prices are determined using ex-post LMP values which in turn rely on the actual SCADA measurements to calculate their final settlement prices.  
Thus, if an attacker can manipulate the system measurement data, the results of the SE, and consequently, the electric energy price can be affected. The authors use a linear form of Optimal Power Flow (OPF), DCOPF, to calculate the LMPs and formulate the attack as a convex optimization problem. There are two cases applied to the IEEE $14$ bus system, one for a single congested line and the other for three congested lines.} The study illustrates that FDIAs can manipulate the nodal price of the ex-post market and can also bring financial profits to attackers. The authors also explore, in a later study~\cite{xie2011integrity}, more realistic attack scenarios assuming threat models in which the attackers can only manipulate a limited number of sensors.

\section{Defenses Against False Data Injection Attacks}\label{sec4}

\begin{table}[t]
\caption{Defense strategies against FDIAs.}
\label{tab16}
    \centering
    \begin{tabular}{|c|c|c|}
    \hline
    \textbf{Methods}  & \textbf{References} & \textbf{Limitations}\\
\hline
\makecell{Protecting minimum \\sets of meters} & ~\cite{bobba2010detecting, dan2010stealth, kosut2011malicious, bi2014graphical, mishra2017price}&\makecell{ Protected only\\ measurements \\that are trusted}   \\ 
\hline  
\makecell{Using PMUs} & ~\cite{chen2006placement, kim2011strategic}&\makecell{Vulnerable to \\GPS spoofing \\attack} \\
\hline
\makecell{Defenses based on\\ game theory}& ~\cite{wei2018stochastic, ma2013markov, wang2017determination, sanjab2016data}& \makecell{Rationality of \\the agents and \\modeling \\challenges}\\
\hline
\makecell{Defenses based on \\cryptographic methods} & ~\cite{sun2012dynamic, manandhar2014combating, abdallah2016efficient}& \makecell{Not practical \\for large systems \\with
a limited \\budget} \\ 
\hline
Topology defence method & ~\cite{shahid2018proposed}&\makecell{It is possible for \\ the attacker to \\learn and guess \\the new \\configuration}\\
\hline
\makecell{Proactive approaches \\to mitigate FDIAs} & ~\cite{li2018pama}& \makecell{Computationally \\intensive} \\ 
\hline
    \end{tabular}
\end{table}

In this section, we discuss existing countermeasure approaches against FDIAs. Table~\ref{tab16} lists different detection methods and their limitations. 

Liu \textit{et al}. show that if the attacker knows the system matrix $\mathbf{H}$ and can compromise $k \geq m-n+1$ meters, then she/he can effectively inject the malicious vector to the measurement vector $\mathbf{z}$ without being detected \cite{liu2011false}. Therefore, it is crucial to identify and protect a set of meter measurements. Bobba \textit{et al}. highlight the requirement to identify and protect a set of measurements to prevent FDIAs~\cite{bobba2010detecting}. Both studies leverage a brute force approach to identify the set of measurements that require protection. Dan \textit{et al}. propose a greedy algorithm to find the minimum set of measurements essential to be protected~\cite{dan2010stealth}. Due to the large number of meters in the power system and the limited protection budget, the authors consider protecting a subset of meters $\rho$ to increase the security level of the system. Subsequently, the authors consider two objective functions:
\begin{enumerate}[leftmargin=3em,labelsep=1em]
    \item Maximize the minimum attack cost:
    \begin{equation}
    \begin{split}
            &\rho^{m} = \max_\rho \:\:\min_k \alpha_k\\
            &\mbox{Subject to}: \varrho(\rho)\leq \pi
        \end{split}
    \end{equation}

where $\varrho(\rho)$ is the protection cost, $\pi$ is the budget, and $\alpha_k$ is defined as the security index of the $k$-th measurement (Eq. \eqref{75}).

\item Maximize the average attack cost: 
    \begin{equation}
    \rho^m = \max_\rho \:\:  \frac{1}{m}\sum_{k\in M}\alpha_k
    \end{equation}
\end{enumerate}

In order to minimize the protection cost, Bi \textit{et al}. frame the protection problem as a variant of the Steiner tree problem in a graph \cite{bi2014graphical}. Given an undirected graph with non-negative edges and a set of vertices which represent transmission lines and buses in the power network, the Steiner tree problem entails finding a tree with minimum weights which contains all the vertices \cite{ansari2018graph}. To select the minimum set of meters to be protected, they propose two algorithms: a Steiner vertex enumeration algorithm and MILP. The proposed algorithms significantly reduce the computational complexity and are able to find the minimum set of meters necessary to be protected. The shortcomings of protecting a minimum set of meters are twofold,  \textit{(i)} possible decrease of redundancy,  and \textit{(ii)} occasional lack of security. 

Kim \textit{et al}. propose another approach to protect the minimum set of measurements ~\cite{kim2011strategic}. The authors suggest installing PMUs in the critical substations of electric power systems. PMU is a GPS-based measurement device that directly measures synchronized voltages and phase angles. The GPS connected to the PMU devices time-stamps the measurements, thereby preventing the measurements from being compromised by attackers. Even though installing PMUs is a powerful solution to prevent FDIAs, it is costly to deploy PMUs on a large scale. The cost of a large scale PMU deployment has led to additional research on the optimal placement of PMUs in power systems. To reduce the number of PMUs used in a system, Chen \textit{et al}. develop a placement algorithm to find out locations for PMU installations ~\cite{chen2006placement}. In addition to cost concerns, PMUs are vulnerable to GPS spoofing attacks which could invalidate the PMU data time-stamps (by faking the GPS signal) and compromise the reliability of all the synchrophasor data ~\cite{konstantinou2017gps}.

{In terms of approaches that utilize game theory concepts, Wei \textit{et al}. propose a stochastic-based approach for the protection of the power system from coordinated attacks~\cite{wei2018stochastic}.} Coordinated FDIAs manipulate power system measurements -- by emulating the real behavior of the system -- and thus remain undetectable. The authors design an optimal load shedding algorithm to assess the effects of coordinated attacks, e.g., where and how many loads to be shed under successful attacks. The effect of the attack is then used in a resource allocation stochastic game, to model the interactions between a malicious attacker and a defender. The authors prove the effectiveness of the proposed approach in protecting the power system from FDIAs. However, the game theory model is not scalable, nor realistic, since it models the interaction between defenders and attackers as a series of causal events ~\cite{liang2012game}.

Sun \textit{et al}. propose an encryption-based method leveraging a dynamic secret to protect wireless communications \cite{sun2012dynamic}. The method encrypts the measurement data by using the aforementioned secrets which are dynamically generated at the sender side to protect the security and privacy of the power data. To create the encryption key, instead of using the transmitted data which are vulnerable to eavesdropping attacks, the authors utilize a packet re-transmission communication protocol. The re-transmission protocol employs steganography to encrypt the measurement data rather than send them in plaintext.  {Although encryption methods can protect measurements against FDIAs, they introduce computation overheads and increase communication latency which may become impractical for large and densely interconnected systems with limited edge-computing resources. In order to address computation overheads induced by encrypted communications (for the measurement exchange), the authors in \cite{zografopoulos2020derauth} propose a lightweight hardware-based security primitive which leverages real-time battery entropy for ephemeral key generation and secure authentication between power system assets. 
}

Shahid \textit{et al}. propose a new topology defense model to protect the power system from stealthy blind FDIAs \cite{shahid2018proposed}. The authors exploit the concept of dummy measurement values in the power network to detect stealthy attacks in the network. In their model, meters in the smart grid send two different sets of measurements to EMS  which include dummy and real measurements. The dummy measurements rely on the real measurements and are assigned by operators at the control center. The dummy measurements are only known to the SO. Thus, the SO could quickly detect any attack in the system by comparing all the received measurements against the dummy measurements. However, the mentioned defense model can protect the system only for a limited duration since the attacker can eventually learn or guess the configuration.

Beibei \textit{et al}. introduce a Proactive Approach to Mitigate FDIAs (PAMA) in smart grids \cite{li2018pama}. PAMA can protect the crucial system information such as the original measurement data, system configurations, and grid connections from leakage or even theft. The proposed method focuses on the proactive prevention and mitigation of FDIAs before an attack is conducted. To enhance system robustness against FDIAs, the authors design a distributed computing model that integrates the Paillier cryptosystem to encrypt all system information (including the original measurement data, system configurations, and grid connections). However, PAMA is computationally intensive and challenging to model.

\section{Machine Learning for FDIAs Detection}\label{sec5}

\begin{table}[t]
\caption{Summary of machine learning methods to detect FDIAs.}
\label{tab17}
\centering
\begin{tabular}{|c|c|c|}
\hline
\makecell{\textbf{Type of Machine} \\ \textbf{Learning Methods}} & \textbf{Algorithms} & \textbf{References}  \\
\hline
\makecell{Supervised learning} & \makecell {Support vector \\machine and KNN} &\makecell { ~\cite{ozay2016machine, svm2}\\{~\cite{Detecting11, Ayad2011}}}\\ 
\hline
\makecell{Semi-supervised\\ learning} & \makecell {Semi-support \\vector machine} & \makecell {~\cite{foroutan,ozay2016machine, joachims1998making}} \\
\hline
\makecell{Unsupervised \\learning} & \makecell{Fuzzy $c$-mean} & \makecell{~\cite{mohammadpourfard, yang_false_2018, mohammadpourfard2017statistical}} \\
\hline
\makecell{Deep learning} &\makecell {Multilayer \\Perceptron (MLP)\\\makecell{Recurrent Neural\\ Network (RNN)} \\Deep Belief Network\\ (DBN)} &\makecell{~\cite{foroutan, mohammadpourfard2017statistical, ganjkhani2019novel, tabakhpour2019neural}\\~\cite{Ayad2011,basumallik2019packet, james2018online, he2017real, wei2018false}}\\
\hline
\end{tabular}
\end{table}

Machine learning is a form of artificial intelligence that enables computers to learn and improve without being explicitly programmed \cite{schuld2015introduction}. Different machine learning algorithms have been proposed by researchers to enable FDIAs detection. The existing use of machine learning algorithms can be categorized as shown in Table~\ref{tab17}. In this section, we first discuss the metrics used for the performance evaluation of machine learning -based FDIAs detection algorithms. Leveraging these metrics, we review existing literature and compare their performance.

\subsection{Performance Metrics}

{A multitude of metrics have been adopted to evaluate the performance of the detection methods. Accuracy, Precision, Recall, F1 score, and Receiver Operating Characteristic (ROC) curve are among the most common metrics. With the true label of a measurement and its predicted label, the output of a detection model can be divided into True Positive (TP): indicating a correct positive prediction, True Negative (TN): a correct negative prediction, False Positive (FP): an incorrect positive prediction, and False Negative (FN): an incorrect negative prediction \cite{abdallah2018security}.}

Accuracy is the ratio of the number of correct predictions to the number of total predictions:
\begin{equation}
    Accuracy = \frac{TP+TN}{TP+TN+FP+FN}
\end{equation}
Accuracy is meaningful when the measurement data is balanced (when the number of positive and negative measurement samples are equal). To evaluate the performance of the detection model with imbalanced data, Precision, Recall and F1 score are often considered. {Precision (also called positive predictive value) describes the capability of a model to identify an attack overall true positive predictions ~\cite{koehrsen_beyond_2018}}. It is represented as the ratio of the correct positive predictions to the number of samples labeled as positive: 
\begin{equation} \label{eq:1}
    Precision = \frac{TP}{TP+FP}
\end{equation}

{The Recall (also called sensitivity) gives the model the capability to identify all attacks \cite{koehrsen_beyond_2018}}. Recall is described as the ratio of the number of correct positive predictions to the number of positive samples:  
\begin{equation}\label{eq:2}
    Recall = \frac{TP}{TP+FN}
\end{equation}
From Eqs. (\ref{eq:1}) and (\ref{eq:2}), it can be observed that Precision and Recall are closely related. For a given model, a decrease in FP (Precision) leads to an increase in FN (Recall), and vice versa. To achieve an optimal trade-off between Precision and Recall, the F1 score is used to combine these two metrics. To avoid being heavily impacted by extreme values of Precision or Recall, F1 score is designed as the harmonic average of Precision and Recall, as shown below:
\begin{equation}\label{eq:f1}
    F1\, score = \frac{2\times (Recall\times Precision)}{Recall+Precision}
\end{equation}

For a given classifier, Recall and Precision may vary a lot with different sets of measurements, e.g., balanced data and unbalanced data, making it hard to evaluate the performance of a classifier. In order to have a stable representation of the classifier performance, the ROC curve is often used. In a continuous binary classifier, the output is a continuous variable ranging from $0$ to $1$. Thus, a threshold is leveraged to divide the outputs into positive and negative. Different thresholds result in different TP Rates (TPR, equals to Recall) and FP Rates (FPR = FP/(FP+TN)). ROC curve plots TPR ($y$-axis) against FPR ($x$-axis) under different thresholds. The closer the ROC curve is to the upper left corner or coordinate ($0$,$1$) (the larger the Area Under the Curve -- AUC), the better the performance. ROC illustrates how well the detection method distinguishes between the attacked and the secured measurements.

\subsection{Supervised Learning Algorithm}
Machine learning algorithms can be classified into supervised learning, semi-supervised learning, and unsupervised learning. In supervised learning, inputs and desired outputs are provided to the machine in order to construct a function that maps the input to the desired output. 

{Detecting FDIAs is considered a supervised binary classification problem. The objective of the binary classifier is to decide whether the given data $\mathbf{s}$ with $m$ features is either $\mathbf{z}$, a normal measurement vector (negative class) or $\mathbf{z_a} =\mathbf{z} + \mathbf{a}$, an attacked measurement vector (positive class) \cite{svm2}}. The output class labels are: 
\begin{equation}\label{85}
   y = \begin{cases}
  \ +1  \text{  for  } a \neq 0\\ -1  \text{  for  } a = 0          
  \end{cases}
\end{equation}
where $\mathbf{a}$ is the attack vector.

The common used supervised learning algorithms are perceptrons \cite{stephen1990perceptron}, Support Vector Machines (SVMs) \cite{cortes1995support}, k-Nearest Neighbors (KNN) \cite{altman1992introduction}, and logistic regression \cite{hosmer2013applied}. In perceptrons, a weight vector $\mathbf{w} \in \mathcal{R}^{M_{Tr}}$ is trained such that the output label, $y_i$, of a sample $s_i$ is predicted by the following classification function:
\begin{equation}\label{perceptron}
   f(s_i) = sign(\mathbf{w} \cdot s_i) = \begin{cases}
  \ +1  \text{  for  } \mathbf{w} \cdot s_i \ge 0 (a \neq 0) \\ 
   -1  \text{  for  } \mathbf{w} \cdot s_i < 0 (a = 0)          
  \end{cases}
\end{equation}
During the training phase, the weight vector is updated for each training sample as $\mathbf{w}(i + 1) = \mathbf{w}(i) + \Delta{ \mathbf{w}}$, where, $\Delta {\mathbf{w}} = \gamma (y_i - f(s_i))s_i$, and $\gamma$ is the learning rate. From the classification function, we can see that the convergence of the perceptron algorithm can be guaranteed when the samples are linearly separable. Therefore, it is suitable for FDIAs detection only when a hyperplane can separate the measurements.

\begin{figure}[!t]
\centering{\includegraphics[width=70mm,height= 35mm,keepaspectratio]{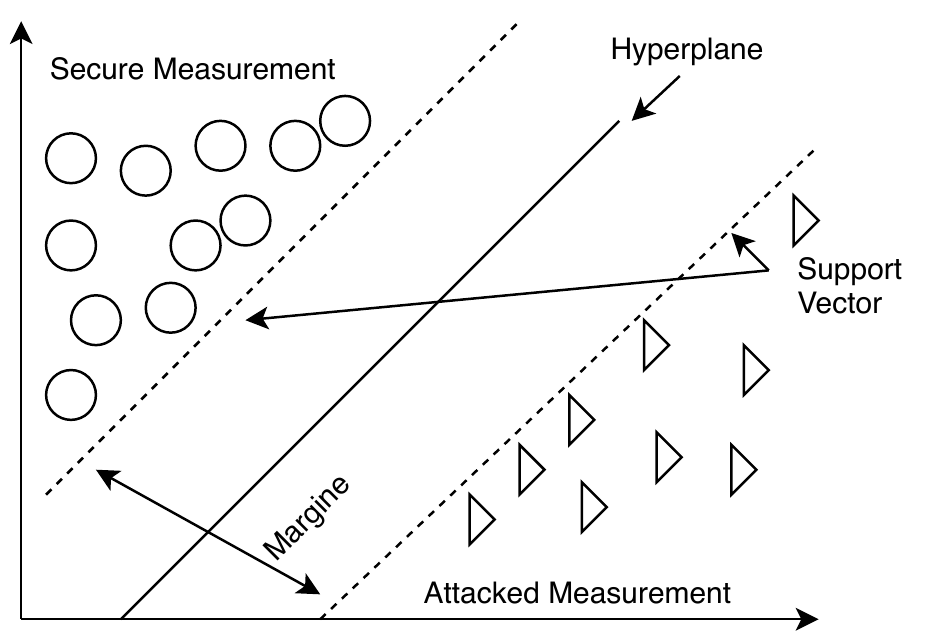}}
\caption{Support Vector Machine concept.\label{sv2}}
\end{figure}

In SVMs, a hyperplane is constructed to separate two different classes. The hyperplane can be represented by a weight vector $\mathbf{w}$, and a bias value $b$. The decision boundaries for the linear separable data can be formulated as two parallel hyperplanes using Eq. (\ref{30}).

\begin{equation}\label{30}
      \begin{cases} \mathbf{w}^Ts_i+b=+1,\text{if  } y_i=+1\\
      \mathbf{w}^Ts_i+b=-1,\text{if } y_i=-1
      \end{cases}
\end{equation}
In Eq. (\ref{30}), each line represents a support vector, as shown in Fig \ref{sv2}. Margin D is the separation area between the two support vectors and can be computed as:
\begin{equation}
D = \frac{2}{\mathbf{w}^2}
\end{equation}
The hyperplanes can be determined by solving:
\begin{equation}\label{svm}
\begin{split}
    &\min_{\mathbf{w},\xi,b} \:\parallel w\parallel^2_2+\zeta\:\sum^{M_{Tr}}_{i=1}\xi_i\\
   &\mbox{Subject to}: \quad  y_i(\mathbf{w}^T\cdot s_i+b)-1+\xi_i\geq 0\\
   & \xi_i\geq 0\:\:\forall i= 1,2,3\dots,\mathbf{M_{Tr}}
\end{split}
\end{equation}
where $\zeta$ is the adjustable regularization parameter, $\xi_i$ is the slack variable for the nonlinear separable training set, and $\mathbf{M_{Tr}}$ is the feature vector. 

KNN is another supervised learning algorithm that assigns labels to an unlabeled sample according to its $k$-nearest neighbors. The Euclidean distance is used to determine the similarity between a given labeled sample, $s_i$, and an unlabeled sample, $s_i'$. The set of KNNs for a given measurement sample can be determined using the Euclidean distance as follows \cite{wang2009divergence}:
\begin{equation}
\begin{split}
&\parallel s_i' - s_{i(1)}\parallel_{2}\:\leq\:\parallel s_i' - s_{i(2)}\parallel_{2}\:\leq\dots\leq\: \parallel s_i' - s_{i(\mathbf{M}_{Tr})}\parallel_{2},\\
& \aleph (s_i')=\{s_{i(1)},s_{i(2)},\cdots\: s_{i(k)}\}
\end{split}
\end{equation}
Majority voting is one of the most commonly used methods for assigning labels from the set of $k$-nearest neighbors of $s_i$. KNN is easy to implement but it fails to work when the size of the data sample is smaller than the dimension of the feature vector \cite{abe2010feature}.

The logistic regression algorithm assumes that the distribution of the label $y_i$ of data $s_i$ follows the following logistic function \cite{cramer2002origins}:
\begin{equation}\label{lr}
P\left(y_{i} | {s}_{i}\right)=\frac{1}{1+\exp \left(-y_{i}\left({\mathbf{w}} \cdot {s}_{i}+b\right)\right)}
\end{equation}
The weight vector $\mathbf{w}$ is estimated by maximizing the following cost function:
\begin{equation}
\mathbf{J}(\mathbf{w})=-\frac{1}{\mathbf{M}_{Tr}}\sum_{i=1}^{\mathbf{M}_{Tr}}\log(1+\exp(-y_{i}\left({\mathbf{w}} \cdot {s}_{i}+b\right)))
\end{equation}

A brief comparison of various machine learning methods is presented in \cite{ozay2016machine}. The paper is one of the first research works to utilize supervised learning algorithms for FDIAs detection. The authors used a hierarchical network in which the measurements are grouped as clusters, and each cluster is regarded as a sample $s_i$. The false data is directly injected into the measurements before the measurements are grouped into clusters. The detection method is based on the observations made in \cite{liu2014detecting}. According to \cite{liu2014detecting}, the distance between samples determines the attack vector:
\begin{equation}
    \parallel s_i - s_j\parallel_2 = \begin{cases} 
    \parallel \mathbf{z}_i -\mathbf{z}_j + \mathbf{a}_i -\mathbf{a}_j\parallel_2,~~\text{if}~~\mathbf{a}_i,\mathbf{a}_j \neq 0\\
    \parallel \mathbf{z}_i -\mathbf{z}_j + \mathbf{a}_i    \parallel_2 ,~~\text{if}~~\mathbf{a}_i \neq 0, \mathbf{a}_j=0\\
    \parallel \mathbf{z}_i -\mathbf{z}_j\parallel_2,~~\text{if}~~\mathbf{a}_i,\mathbf{a}_j = 0
     \end{cases}
\end{equation}
Therefore, by looking into the distance between two samples, it is possible to detect a FDIA. In their experiment, the performance of different machine learning algorithms is evaluated against FDIAs with different sparsity $k/m$ (the ratio of measurements that the attacker has access to). Accuracy, Precision, and Recall are used as performance metrics. The results proved that the machine learning algorithms perform better than any other algorithm (e.g., state vector estimation approach) in detecting FDIAs. Although SVMs achieved the highest prediction accuracy, they also present some limitations, such as the selection of the kernel and sensitivity to the sparsity of the system. {KNN is very sensitive to system size and performed better for the small-sized systems.} Despite conducting plenty of experiments, Ozay \textit{et al}. did not evaluate the performance of detection algorithms for stealthy FDIAs. 
Furthermore, only the sparsity of injected data was considered. The magnitude of the injected data could potentially impact as well as the operation of the system and the performance of the detection methods.  Last, the lack of attack data can result in imbalanced data samples during the detector training process affecting its classification accuracy. 

Considering the limitations of \cite{ozay2016machine}, a similar work is conducted in \cite{svm2}. {Both works utilize closely related system models. Two assumptions are taken into consideration when the attack vectors are created in the adversary model: \textit{(i)} that the injected value $a_i$ is greater than the noise level, and \textit{(ii)} that the mean of the attack vector $\mathbf{a}_i$ is larger than the variance of the attack vector. Attack vectors with different sparsity and variance\footnote{The variance reflects the magnitude of disturbances caused by false data.} are tested in their experiments. To solve the imbalanced data problem, they propose the Extended Nearest Neighbor (ENN) algorithm.} For each class, ENN measures the average ratio of the nearest neighbors belonging to the same class. Instead of using majority voting, the label of a sample was predicted by finding the class which presents the greatest ratio variability with the sample labeled in that class. The performance of SVMs, KNN, and ENN are then experimentally evaluated. Accuracy and F1 scores are used as the performance metrics. SVMs outperformed KNN and ENN in most of the test cases. A critical range of sparsity was observed in which the Accuracy and F1 score increased significantly. However, this is reasonable since the distance increases $\parallel s_i - s_j \parallel$ when the sparsity increases which leads to more distinct classes. The experiment was conducted on the IEEE $30$ bus system. The detection performance of the algorithms in larger systems was not demonstrated. 

Esmalifalak \textit{et al}. propose a distributed SVM algorithm \cite{Detecting11}. Each substation owned a training set and stealthy FDIAs, which could bypass BDD methods based on their corresponding residuals. Before training, PCA is applied to the training set to reduce the feature dimension. To avoid a huge volume of data exchange, each substation is trained using a local classifier, and only the locally optimized weight vectors are exchanged. Their optimization problem, Eq. (\ref{svm}), is provided below:
\begin{equation}
\begin{split}
    &\min_{w_k,\xi_k,b_k}\:\parallel w_{k}\parallel^{2}_{2}+\zeta\:\sum^{n}_{k=1}\sum^{M_{Tr}}_{i=1}\xi_{ki}\\
  & \mbox{Subject to}:\quad y_{ki}(w_{k}^{Tr} \cdot s_{ki}+b_k)-1+\xi_{ki}\geq 0\\
   & \xi_{ki}\geq 0\:\:\forall i= 1,2,3,\dots,\mathbf{M_{Tr}};k=1,\dots,n
\end{split}
\end{equation}
where $n$ is the number of substations and $w_k$ is the local optimization parameter. The Alternating Direction Method of Multipliers (ADMM)  is used to solve this distributed optimization problem. Experiments are performed on the IEEE 118 bus system. The authors empirically verify the convergence of distributed SVM classifiers to centralized SVMs with different numbers of substations. 

{To recapitulate, supervised learning methods have achieved superior performance in comparison to traditional residual-based BDD methods. Among the aforementioned algorithms, SVMs has demonstrated to achieve the highest accuracy. According to ~\cite{Detecting11}, the curse of dimensionality problem can be solved leveraging PCA which significantly enhances the efficiency of machine learning algorithms. Nevertheless, most of the attack data are often generated randomly in the experiments while a sophisticated adversary would deliberately choose attack vectors considering the system dynamics. The performance of the proposed methods against such sophisticated FDIAs still remains unknown. Moreover, most of the prior works conducted simulation experiments. Thus, the efficiency of existing methods, if applied to real power system deployments, cannot be guaranteed.}

\subsection{Semi-supervised Learning} 

In semi-supervised learning, the majority of the given data is unlabeled. Although semi-supervised learning algorithms are the least common learning approaches applied for detection of FDIAs, we still introduce them in this sruvey work for completeness. An example of a semi-supervised learning algorithm is a semi-supervised SVM or $S_3VM$. $S_3VM$ assumes that samples with different labels are clustered into different groups, and that the diameter of each cluster is small enough to avoid sub-clusters \cite{chapelle2008optimization}. The objective function of $S_3VM$ is defined as:
\begin{equation}\label{eq:s3vm_}
\min_{\mathbf{w},b}\: \zeta\Big[\sum_{i=1}^{\mathbf{M}_{Tr}} L^{Tr}(s_i,y_i) +\sum_{i=1}^{\mathbf{M}_{Ts}} L^{Ts}(s'_i)\Big] +\parallel w\parallel^2
\end{equation}
where $y=w^Ts_i+b$ and $\zeta$ is the regularization parameter, $L^{Tr}$ and $L^{Ts}$ are the loss function of the training and test samples,  respectively.

Foroutan \textit{et al}. investigate FDIA detection methods by using the $S_3VM$ based on Gaussian mixture distributions~\cite{foroutan}. According to \cite{costa2008mixture}, a finite mixture distribution model, defined as a convex combination of two or more probability density functions, is capable of approximating any arbitrary distributions due to its flexibility in modeling complex data. The authors assume that all FDIAs have the same amount of energy or $\mathbf{c}$ vector, where $\mathbf{a}=\mathbf{H}\mathbf{c}$, and that they have the same mean squared error. In the adversary model, the attack vectors are designed based on the minimum energy residual attack and sparsest attack, introduced in \cite{kosut2011malicious} and \cite{hendrickx2014efficient}, respectively. In the training phase, a positive dataset, i.e., a dataset with attacked measurements, was used to build the Gaussian mixture model. Then, a mixture of a datasets consisting of both positive and negative labels (attacked and normal measurements)  determines the threshold. In the evaluation phase, the unlabeled dataset used for testing and F1 score evaluates the performance of the results. PCA was applied to the dataset to overcome measurement dimensionality issues. The authors demonstrate the performance of the proposed detection method on the IEEE $118$ bus power system. To generate diversified datasets, different topological networks are constructed using Monte Carlo simulations. The performance of the proposed detection method depends on the selection of a proper threshold. A high threshold value reduces Recall while a low threshold value lowers Precision. The impact of the detection algorithms is illustrated with a ROC curve. Although the proposed model demonstrates a high F1 score compared with other machine learning algorithms (e.g., SVMs and perceptrons), it performs well only when the attacked measurements and the real measurements lie in distinct regions of the feature space, i.e., the attacked data can be effortlessly isolated.

Another detection method based on the ($S_3VM$) algorithm is proposed in \cite{ozay2016machine}. The input samples are integrated into the cost function forming the following optimization problem:
\begin{equation}
    \min \parallel W \parallel_2^2 + \zeta_1 \sum_{i=1}^{M_{Tr}} L^{Tr}(S_i,y_i)+\zeta_2 \sum_{i=1}^{M_{Te}} L^{Te}(S_i^{'})
\end{equation}
where $\zeta_1$ and $\zeta_2$ are the cost parameters, $L^{Tr}$ and $L^{Te}$ are the loss functions for the training and testing samples. In the simulation, the authors use default values for the parameters as suggested in \cite{joachims1998making}. The experiment are conducted on IEEE $9, 57,$ and $118$ bus systems, and the measurement matrix is generated using Matlab's Matpower toolbox. Compared with supervised learning algorithms,  $S_3VM$ demonstrated improved robustness against data sparsity despite the fact that $S_3VM$ still remains sensitive to unbalanced data samples.

\subsection{Unsupervised Learning}
Unsupervised learning algorithms group the unlabeled samples based on the similarities and differences between samples, without any prior training. Clustering is the most popular unsupervised learning method where the measurement samples are grouped based on the distance between samples in the feature space. Different distance metrics can be chosen (e.g., Euclidean distance). 

$k$-means (also called hard $c$-mean) is an example of a clustering method which divides data into $k$ groups. $k$-means iteratively assigns each data point to one of the $k$ groups, whose centroid has the minimum distance to the data point in the feature space. Each centroid in a cluster is a collection of features which define a group. The centroids are updated at each round. Several techniques are used to validate the $k$-value including cross-validation and other information criteria. 
Fuzzy $c$-means clustering (FCM) is another type of $k$-means clustering which assigns data points to two or more clusters \cite{mohammadpourfard}. Each point belongs to a cluster based on a corresponding probability value, rather than having a binary value as is the case of $k$-means clustering. In FCM, the clustering problem can be solved by minimizing the following equation:
\begin{equation}
    J = \sum_{i=1}^N \sum_{j=1}^C u_{ij}^m \parallel x_i - C_j\parallel^2\: ,\: 1\leq m<\infty
\end{equation}
where $N$ is the number of data points, $C=2$ is the number of the clusters (cluster of attacks and cluster of normal measurements), $x_i$ is the $i$-th dimensional measured data, and $C_j$ is the center of the $j$-th cluster, which is determined using:
\begin{equation}
    C_j= \frac{\sum_{i=1}^N u_{ij}^m \cdot x_i}{\sum_{i=1}^N u_{ij}^m}
\end{equation}
where $u_{ij}$ is the degree of membership of the $i$-th measurement. The updated membership $u_{ij}$ computed by the following equation: 
\begin{equation}
    u_{ij}=\frac{1}{ \sum_{k=1}^C\bigg(\frac{\parallel x_i-C_j\parallel}{\parallel x_i-C_k\parallel}\bigg)^{\frac{2}{m-1}}}
\end{equation}

Mohammadpourfard \textit{et al}. present a visualization based on the unsupervised anomaly detection method and FCM clustering to detect and locate FDIAs \cite{mohammadpourfard2017statistical}. The authors also propose a localization method that helps in identifying the attack after topology reconfigurations and the integration of different resources. When FDIAs occur, the probability distributions of system states deviate significantly from normal states, hence enabling FDIA detection. First, the authors normalize the data, and then various statistical measures are applied to characterize the probability distribution of each state vector. PCA is applied to the new feature set to reduce the dimensionality of data and to visualize them in a two-dimensional space where the grid operators can determine whether an attack has occurred or not (using patterns of normal and abnormal data). FCM is used to detect outliers and locate the FDIAs. Load data from the New York Independent System Operator (NYISO) are used for the simulations. FDIAs data are generated on the IEEE $9$, and $14$ bus system with the assumption that the adversary decreases or increases a specific state variable by at least $6\%$ of its original value. The proposed method is applied to two different FDIAs scenarios: detecting FDIAs with and without topology changes. Compared to supervised learning algorithms such as SVMs and KNN, the proposed model achieves higher detection accuracy~\cite{mohammadpourfard2017statistical}.

Yang \textit{et al}. proposed three different anomaly detection approaches to detect FDIAs: \textit{(i)} local outlier factor, \textit{(ii)} isolation forest, and \textit{(iii)} robust covariance estimation \cite{yang_false_2018}. The local outlier factor is a density-based anomaly detection method that measures the local standard deviation of any given data point from its neighbors by comparing their local density \cite{8894484}. Isolation forest is an outlier detection technique based on decision trees that does not employ any distance or density measure and can handle large, high-dimensional datasets. Robust covariance estimation is another anomaly detection method based on the elliptic envelope fitting method, which assumes that the given data is a Gaussian distribution and defines the shape of the data. An IEEE $14$ bus system case is used to evaluate the mentioned detection approaches. Attack vectors generated with Gaussian distributed non-zero elements have the same mean and variance as the original measurement set. The authors use PCA to reduce the data dimension from $41$ to $2$,  to reduced noise, and simplify the detection problem. All proposed methods achieve high accuracy for FDIAs detection. However, these three detection methods achieve high detection rates only when the contamination rate is known and small~\cite{yang_false_2018}.

\subsection{Deep Neural Network}

Deep learning algorithms mimic the human brain structure, functions, and are one of the fastest developing artificial intelligence technology. Although deep learning algorithms require time and large amounts of data for their training stage, they have been applied for FDIAs detection achieving high accuracy rates \cite{hassabis2017neuroscience}. 

\begin{figure}[!t]
\centering{\includegraphics[width=80mm,height= 35mm,keepaspectratio]{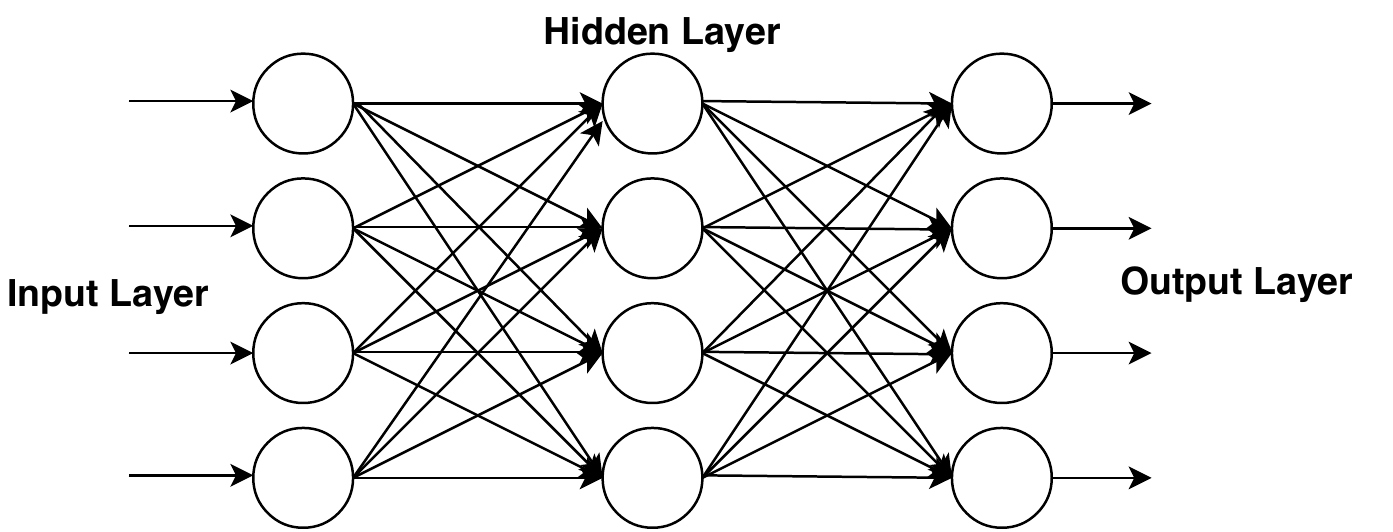}}
\caption{Multilayer Perceptron concept.\label{mlpr}}
\end{figure}

Multilayer Perceptrons (MLPs), also called feed-forward neural networks, are deep learning models where information flows in only one direction, i.e., from the input through the hidden layers to the output, as shown in Fig. \ref{mlpr} \cite{ujjwalkarn_quick_2016}. They consist of an input layer which receives the input signals, one or more hidden layers to construct the approximation function, and an output layer that predicts the final decision based on the input and the approximation function.

Multiple studies where MLPs have been applied to detect FDIAs have been reported in literature \cite{ mohammadpourfard2017statistical, ganjkhani2019novel, tabakhpour2019neural}. In these works, the FDIA detection problem is formulated as a supervised classification problem. In MPL-assisted binary classification, a linear combination of an input weight vector produces a single output, as shown in the following equation:
\begin{equation}
    y= \varphi (\sum_{i=1}^n w_is_i+b)
\end{equation}
where $y$ is the estimated output of the activation function, $w$ is the weight, $\mathbf{s}$ is the input vector, $b$ is the bias, and $\varphi$ is the nonlinear activation function. The activation function is an essential feature of the MLP architectures. It decides whether a neuron should be fired or not by calculating the weighted sum of inputs and adding a corresponding bias to it. Sigmoid, Tanh, and Rectified Linear Unit (RELU) are examples of activation functions. RELU is the most widely used function because it is fast and less computationally expensive. MLPs use back-propagation training algorithms, and the weights are updated using gradient descent to minimize the error function.

Ashrafuzzaman \textit{et al}. propose different MLP structures for the detection of FDIAs in an AC static SE system topology \cite{ashrafuzzaman2018detecting}. The paper assumes that partial knowledge of the system, including the $\mathbf{H}$ matrix and other parameters, is known to the attacker. A standard IEEE $14$ bus system is used to conduct the simulation. The Matpower toolbox is used to generate the measurement vector $\mathbf{z}$, which contains $122$ measurement features ($40$ active and reactive power flows, $14$ power injections, and $27$ voltage measurements). The authors train the MLP using stochastic gradient descent (an optimization technique for the network parameters update) and tanh as the activation function. Four models with different network architectures are utilized for the detection. The first model consists of one hidden layer with $100$ neurons, and the second model consists of $3$ hidden layers with $150$ neurons. For the third and fourth models, the authors use the first and second models with a regularization value of $0.0001$. Regularization is a technique used to reduce or prevent overfitting of a neural network. The models' detection performance is compared with other machine learning algorithms. Accuracy, Precision, Recall, and F1 score are used to evaluate the MLP detection. The results of the four discussed models are similar with an accuracy of around $98\%$.

{Similarly, Foroutan \textit{et al}. apply MLPs to detect FDIAs and compared them with common machine learning detection models \cite{foroutan}.} The network consists of an input layer, one hidden layer, and an output layer. Tanh is used as an activation function. Although MLP produced higher detection accuracy compared to the other algorithms, their training process is slow. 
Ganjkhani \textit{et al}., on the other hand, introduce a novel MLP algorithm leveraging a Nonlinear Autoregressive Exogenous (NARX)  configuration which takes into account the high correlation between power system measurements as well as the state variables \cite{ganjkhani2019novel}. The NARX configuration is used for time series prediction and can predict step-ahead values of the states by factoring measurement values and historical data as input variables. In the experiment, NARX is constructed with an input and a hidden layer with different numbers of neurons and Sigmoid linear activation functions. The historical data contained $6048$ measurement vectors and state variables. The detection model is trained using $70\%$ of the historical data, and $30\%$ for the testing and validation.

\begin{figure}[!t]
\centering{\includegraphics[width=80mm,height= 35mm,keepaspectratio]{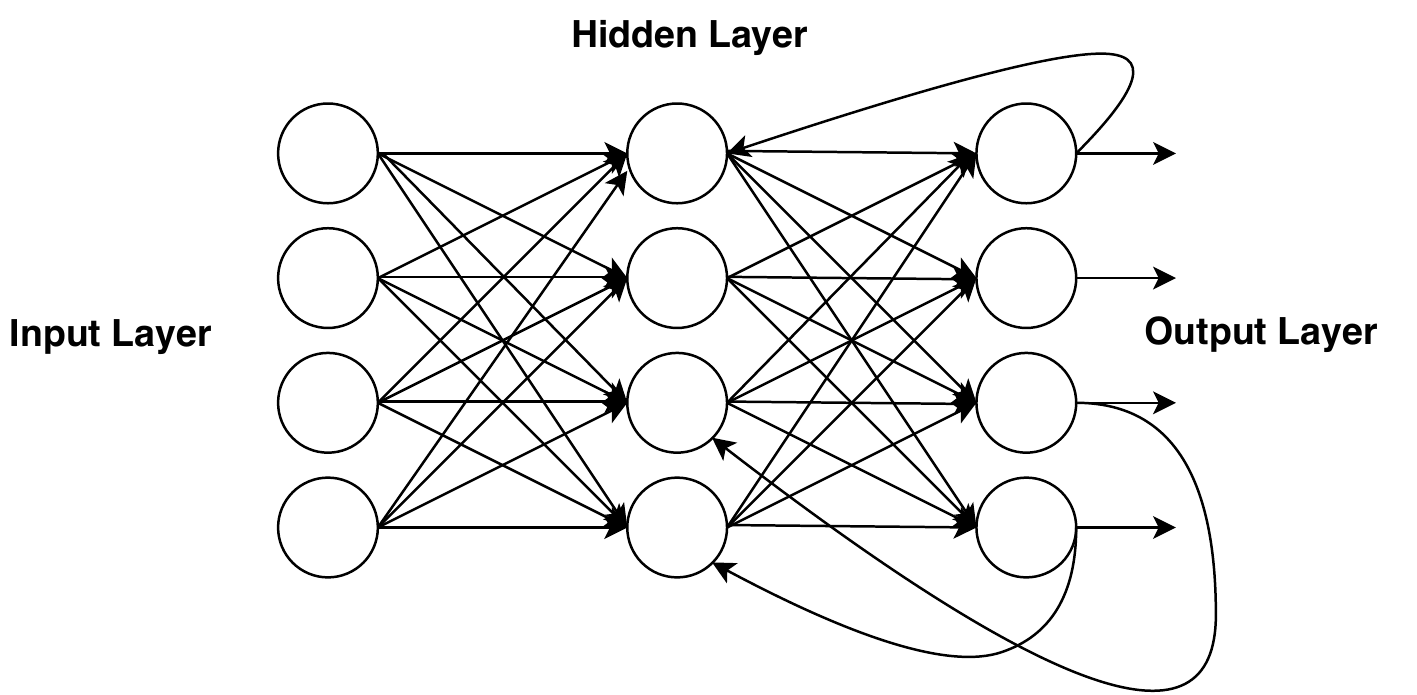}}
\caption{Recurrent Neural Network concept.\label{RNN}}
\end{figure}

A Recurrent Neural Network (RNN) is a sophisticated deep learning algorithm that uses internal memory or feedback loops, as shown in Fig. \ref{RNN}. Unlike MLPs, RNNs use the information from past events for their predictions. A Long Short Term Memory (LSTM) unit can be added to a standard RNN to solve the problem of vanishing gradient descent and store information for an extended period \cite{hochreiter1997long}. RNNs formulate the FDIAs detection problem as a sequence of prediction. Results from previous time steps are used for the prediction of the current output rendering RNNs efficient in detecting manipulated measurements. 

Ayad \textit{et al}. utilize RNNs as a sequence classification algorithm for detecting FDIAs in DC SE \cite{Ayad2011}. Back-propagation through time, an extensive type of back-propagation, is applied to train the algorithm. The authors run the training algorithm multiple times to produce the optimal set of parameters that achieve the least error. Then, the optimal parameters are applied to the network for the prediction of the test data classes. Since the output ranged from $0$ to $1$, a threshold is set to determine the output class as either $1$ or $0$ ($1$ is compromised, and $0$ is normal). IEEE $30$ bus test case is used for the experiments with $112$ measurement vectors. The proposed model obtains outstanding detection results with an accuracy rate of $99\%$.    

Yu \textit{et al}. propose an RNN architecture for detecting FDIAs in AC SE setups \cite{james2018online}. The Discrete Wavelet Transform (DWT) algorithm is used for the RNN model. The main goal of DWT is to extract the hidden time-frequency domain characteristics and features at every specific time. The proposed model is able to leverage dynamic temporal and spatial features for attack detection. The authors detect FDIAs in AC SE with complete and incomplete system knowledge. For the incomplete knowledge case, the attacker only knew a few selected phase angles, power flows, and power injections for selected buses. The remaining buses information is generated using the algebraic sum of the connecting buses. The RNN based detection model is constructed with two types of neuron layers: Gated Recurrent Unit (GRU) and fully-connected dense layers. To tune the network hyperparameters, a dropout approach is used. In dropout, the outputs of some layers are discarded according to predefined probabilities. Dropout solves the overfitting problem, eminent in extensive training datasets and increases the accuracy for newly added test data. The proposed detection method's performance is assessed on IEEE $118$ bus and $300$ bus test cases. Over $200k$ samples are generated to train the detection model, and which achieve a high detection rate of $93\%$ \cite{james2018online}. However, the main challenge of this RNN type is to optimally tune the network hyperparameters.

Deep Belief Networks (DBN) consist of multiple layers of stochastic and latent variables \cite{Hinton:2009}. The latent variables are generally binary variables. DBNs are compositions of simple, unsupervised networks such as Restricted Boltzmann Machines (RBMs) or autoencoders \cite{Hinton:2009}. The authors in \cite{aboelwafa2020machine}, utilize autoencoder networks for the detection of FDIA leveraging temporal and spatial sensor data correlations. He \textit{et al}. propose a DBN and State Vector Estimator (SVE) for real-time detection of FDIAs \cite{he2017real}. The proposed model utilizes an extended DBN called Conditional Deep Belief Network (CDBN) which extracts temporal features in high-dimensions. SVE calculates the $\L_2$-norm of the measurement residuals and compares them with a given threshold as follows \cite{he2017real}:
\begin{equation}
    \begin{cases}
    \mathbf{ r}=\parallel \mathbf{\hat z} -\mathbf{H} \mathbf{\hat x}\parallel_2 \ > \tau,~~\text{Attack alarm}\\
      \mathbf{r}=\parallel \mathbf{\hat z} -\mathbf{H} \mathbf{\hat x}\parallel_2 \: \leq \tau,~~\text{No attack alarm} \\
    \end{cases}
\end{equation}
The authors design the model based on the assumption that the topology of the power system does not change significantly within a small time-frame. For the simulation, the IEEE $118$ bus test case is used to simulate four different attack scenarios. ROC curve is used to evaluate the detection scheme. Then, the detection results with a different number of attacked measurement $k$ are compared with other detection algorithms such as MLPs and SVMs. The proposed model achieves the highest detection accuracy. However, training a DBN is extremely computationally expensive, since this process can take up to weeks even if specialized hardware exploiting GPU acceleration is used \cite{sarikaya2014application}.

Wei \textit{et al}. propose a different DBN-based model~\cite{wei2018false}, where the detection process can be divided into three parts: \textit{(i)} the data pre-processing stage, \textit{(ii)} the training stage, and \textit{(iii)} the testing stage. During the pre-processing data stage, measurement data including the attacked measurements are extracted using different IEEE standard nodes. The training process of the DBN is divided into the pre-training stage, and reverse-trimming stage. In the pre-training stage, the authors use an unsupervised greedy learning algorithm from the bottom layer to the upper layer to extract the measurement features, train every layer, and share the measurement features with every layer. The RBM is trained layer-by-layer and tuned using back-propagation to minimize prediction errors. After the training process, part of the measurement data is used to test and validate the performance of the detection model. The simulation results show that the DBN-based detection achieves high accuracy in detecting FDIAs ($98\%$).

\section{Discussion on the Detection Performance of Machine Learning Algorithms}\label{sec6}
{This section discusses the performance of the machine learning-based FDIA detection algorithms presented in Section~\ref{sec5}.} 
{A noteworthy advantage of such  detection algorithms is that they do not assume exact knowledge of the power system model nor its corresponding parameters, thus any induced uncertainties, e.g., measurement noise, topology changes, power flow perturbations, etc. do not affect the algorithm's detection efficacy.}

{The majority of FDIA detection research focuses on the transmission level, and studies which examine detection algorithms involving Automatic Generation Control (AGC) and  wind generation have been reported \cite{mohammadpourfard2017identification, beg2017detection, tan2017modeling}. On the other hand, studies that examine FDIAs detection for DSs are also essential and a direction of ongoing research \cite{deng2018false}.}
{Our investigation suggests that FDIAs studies can be broadly classified under two major categories, \textit{(i)} random FDIAs, where the attacker aims to inject falsified attack vectors and compromise the SE algorithm by modifying any measurement vector that can be attained, and \textit{(ii)} targeted FDIAs, in which the attacker objective is to inject specific errors into the SE algorithms by maliciously modifying distinct measurement vectors. Apart from the aforementioned FDIAs types, studies involving stealthy FDIA detection have also been proposed \cite{Detecting11}.}

{The detection accuracy of the machine learning FDIA algorithms yields significantly different results depending on the setup used to evaluate the algorithm's performance. For example, some studies -- in order to characterize the detection performance -- examine algorithms under hundreds of different FDIA scenarios and varying power system topologies, while others report results based on very limited datasets. Notably, the algorithms presented in \cite{ozay2016machine, mohammadpourfard2017statistical, james2018online} are thoroughly tested on multiple power system architectures, such as IEEE $14, 30,$ and $118$ bus systems, contrary to the algorithms in \cite{teixeira2011cyber, foroutan, yang_false_2018},which utilize only one IEEE system model during the performance analysis. Additionally, some papers consider basic FDIAs while others evaluate detection performance against stealthy FDIAs, which significantly skews the algorithm efficacy \cite{Detecting11, ashraf}. Finally, a number of researchers develop their custom metrics to assess the proposed detection algorithms, or do not provide any quantitative results whatsoever. For all the aforementioned reasons, providing an overarching algorithm comparison or declaring an optimal detection algorithm for every case is infeasible, since detection performance is contingent upon a multitude of reasons (e.g., TS or DS, stealthy or basic FDIAs, size of the system under test, etc.) and comprehensive results are not available in the literature.}

{SVMs are consistently more effective in detecting FDIAs in power systems with reported detection rates ranging from $85\%$ to $99\%$ \cite{ozay2016machine, svm2, Detecting11}. Contrary to supervised learning detection methods, SVMs do not require exhaustive training and big data sets which increase computational complexity and training duration~\cite{zhou2018brief}. On the other hand, SVMs performance can degrade significantly if the kernel selection process is not properly conducted or when we deal with sparse systems~\cite{nayak2015comprehensive, chen2009mining}. Another drawback of SVMs,  which has been recently reported and can effectively lower their detection accuracy, is that their susceptibility to adversarial examples~\cite{sayghe2020adversarial,}. Adversarial examples are carefully crafted inputs intentionally designed to falsify machine learning algorithms~\cite{goodfellow2014explaining}. For instance, label flipped attacks are a form of adversarial example which targets SVMs and affect their detection competency against FDIAs in power systems~\cite{sayghe2020adversarial, saygheevasion}.}

{Apart from SVMs, deep learning algorithms have been proposed for the detection of different types of FDIAs (e.g., stealthy or basic) and in different power system topologies (i.e., TS or DS). Contrary to SVMs, deep learning methodologies require large amounts of training data and their detection efficiency is heavily affected by the dimension of the training dataset. Multiple works report detection rates between $90\%$ and $99\%$ when abundance of training data is available for the deep learning detectors~\cite{foroutan, mohammadpourfard2017statistical, ganjkhani2019novel, tabakhpour2019neural, Ayad2011,basumallik2019packet, james2018online}. Despite the impressive results that deep learning algorithms exhibit, their training process requires an excessive amount of time, has high computational costs and demands specialized equipment, in addition to big datasets. For instance, the authors in~\cite{he2017real} report that more than $3k$ measurement samples are essential in order for their deep learning algorithm to achieve detection rates of $98\%$.}

{Previous works prove that machine learning algorithms including supervised learning, SVMs, and deep learning method  are able to effectively and in real-time detect FDIAs in power systems~\cite{he2017real, 8894484}. The main pitfall of machine learning approaches is that they require extensive datasets and historical data including attack scenarios to train the detectors~\cite{al2015efficient}, which causes all the aforementioned disadvantages~\cite{ozay2016machine}. Besides the exploitation of resources, e.g., memory, storage space, specialized hardware, etc., overfitting is another vulnerability that machine learning algorithms suffer from. By overfitting a machine learning algorithm we end up with a detector that is able to perform exceptionally well for specific datasets, but cannot generalize this performance for all possible test cases, thus even selecting a proper training set becomes challenging~\cite{dietterich1995overfitting}. Adversarial examples can also compromise machine learning-based algorithms. Limited research works address this issue~\cite{sayghe2020adversarial, tian2019adaptive, chen2018machine, saygheevasion}, thus developing robust machine learning detectors against adversarial examples is imperative and one of our future directions.}

\section{Conclusions and Future Directions}\label{sec7}

{Improving the cybersecurity of cyberphysical energy systems is vital for the efficient and resilient operation of the power grid. FDIAs can elicit severe physical and economic impacts on power systems. Researchers have thoroughly investigated FDIAs and have proposed algorithms to detect these data integrity attacks. Among these algorithms, machine learning-based methodologies are gaining attention due to their superior detection performance.}

{In this paper, we provide a comprehensive review of various FDIA detection methods leveraging machine learning algorithms. The goal of this survey is to compare different machine learning FDIA detectors employed in power systems. Our investigation concludes that supervised learning and deep learning methods achieve the highest detection rates.} 
{Our future work will explore how machine learning-based FDIA detectors perform in DS which incorporate DERs (e.g., microgrids) and what modifications are essential. Also, we aim to develop detection algorithms leveraging Generative Adversarial Networks (GANs) to further improve FDIAs  detection performance against stealthy and more sophisticated attacks.}

\bibliographystyle{IEEEtran} 
\bibliography{references}

\end{document}